%% file: MainText.tex
\documentclass[journal]{IEEEtran}

\usepackage{amsmath,amssymb,amsfonts}
\usepackage[T1]{fontenc}
\usepackage[utf8]{inputenc}
\usepackage{graphicx}
\usepackage[export]{adjustbox}
\usepackage{booktabs}
\usepackage{multirow}
\usepackage{array}
\usepackage{float}
\usepackage{cite}
\usepackage{url}
\usepackage[unicode]{hyperref}
\graphicspath{{mathpix_assets/}}
\DeclareUnicodeCharacter{2713}{\checkmark}
\DeclareUnicodeCharacter{00D7}{\ensuremath{\times}}

\title{An Improved Deep Reinforcement Learning Control Strategy for Traction Dual Rectifiers in EMUs}
\author{Zhigang Liu, Mingwei Tang, Xiangyu Meng, Hui Wang, Qiao Zhang, Haoyu Wang, Mengru Li}

\begin{document}
\maketitle

\input{body_mathpix.tex}

\end{document}

%% file: body_mathpix.tex
\begin{abstract}
Due to the use of PI-based $\boldsymbol{d} \boldsymbol{q}$ current decoupling in the pulse rectifier of CRH5 high-speed trains, the PI parameters directly affect the traction system's control performance. Linearized control may have issues with reference trajectory changes or model mismatches, leading to a decrease in system performance, while nonlinear control may have problems with jitter and poor steady-state accuracy. This paper proposes a new control strategy that replaces all PI in the $d q$ current decoupling control with a single intelligent agent. This method based on Deep Reinforcement Learning (DRL) can avoid various drawbacks of linearization and nonlinear control and ensure the stability of intermediate DC voltage. However, when EMUs are in different working conditions and switching, the Twin Delayed Deep Deterministic Policy Gradient (TD3) algorithm used in traction dual rectifiers does not have a good control effect. Focusing on the issue, Reward Shaping (RS) is added to re-design a nonlinear reward function, which can be combined with Prioritized Experience Replay (PER) to increase the convergence speed of the episode reward. The simulation results show that the improved control strategy can be effectively applied to EMUs working in multiple conditions. Finally, the stability analysis is carried out using Lyapunov's second method and the verification results of the hardware-in-the-loop (HIL) simulation platform show that the DRL control has a good effect.
\end{abstract}

Index Terms-EMUs, dq decoupling control, Deep Reinforcement Learning, traction dual rectifiers, multi-working conditions, hardware-in-the-loop (HIL).

\section*{I. INTRODUCTION}
At present, the power unit of China's EMUs mainly uses AC-DC-AC traction transmission system, which is mainly composed of pulse rectifier, DC circuit, traction inverter, traction motor and so on. All links interact with each other, especially the converters and other power electronic devices have strong nonlinearity and impact, which eventually cause the power quality problem of the coupled system of EMUs and traction network in the different operating conditions of EMUs [1]. Among them, the pulse rectifier needs to maintain the stability of the DC link voltage to provide good conditions for the inverter through the control of the rectifier.

\section*{A. Issue description}
The current pulse rectifier combined with the control strategy adopts sine pulse width modulation (SPWM) technology to control the on/off of IGBT, achieving real-time control of various electrical quantities in the rectifier circuit.

From the development of traditional PID control, modern controls such as variable structure control, robust control, predictive control, etc., and intelligent controls such as fuzzy control, expert control, and neural network control have emerged, which do not rely on the mathematical model of the system. In this paper, the $d q$ decoupling control in CRH5 EMUs is taken as an example to analyze. The double closedloop control structure based on PI has the bandwidth problem of voltage loop and current loop[2], which will affect the stability of the rectifier circuit. At the same time, the selection of multiple PI parameters in the voltage and current loops is related to the dynamic and static performance of the rectifier output. Inappropriate values may exacerbate the negative impedance characteristics of the rectifier and even lead to lowfrequency oscillations [2]. Due to the nonlinear characteristics of converters, the feedback linearizing control [4] and model predictive control [5] need to rely on accurate mathematical models when linearizing the system, which may lead to the deterioration of system performance due to the change of reference trajectory or model mismatch [6]. Since these nonlinear controls are based on energy, there are problems such as chattering and poor steady-state accuracy [7]. The intelligent control of existing converters, such as fuzzy control [8], expert control, and genetic algorithm-based control, does not rely on accurate mathematical models and has good adaptive ability. However, the design of these methods is complicated, the precise analytical solution of the control output cannot be obtained, and the control accuracy and steady-state accuracy are poor.

\section*{B. Literature review}
The application of AI adds a data-driven control strategy to intelligent control, which does not depend on the mathematical model of the system. At the same time, this control strategy uses the strong fitting function of the neural network, which can improve the adaptive ability, and be applied to the system with strong nonlinearity. Most existing AI-based controls adopt the Deep Reinforcement Learning (DRL) algorithm, which can be divided into three categories according to structure: Compensation, Parameter adjustment, and Replacement. The first is the compensation class. The DRLbased compensation control uses the agent's output to compensate the original control's output, for optimizing the original control's performance. However, the main body is still the original control. For example, if the output of PI control is compensated [9], the problem of PI control cannot be completely solved. Another method with the phase adjustment\\[0pt]
of the modulation signal based on DRL [10] can also be classified as compensation because it retains the original control strategy and only uses DRL to adjust the phase of the modulation signal in the SPWM link. It can also be regarded as compensating the output of the original control. The second is the parameter adjustment class. The DRL-based PI control [11] uses the agent's output to adjust PI parameters in real time, and the control performance is better than that when parameters are fixed. However, in the double closedloop control structure with the $d q$ decoupling [2], there are 3 PI controls with 6 parameters that need to be adjusted (after taking into traction dual rectifiers, the parameters increase to 12). Consequently, it is challenging to achieve either the single-agent multi-output control strategy or the multi-agent control strategy. In the replacement class, the original controller is directly substituted with a DRL agent [12], which can also supplant the entire control system and directly produce four control pulses of the two-level full-bridge rectifier [28-30].

\section*{C. Contribution of this work}
\begin{enumerate}
  \item This paper proposes a control strategy that can replace all PI in the $d q$ decoupling control with a single agent, allowing for the successive control of single rectifier and dual rectifiers. This approach simplifies the control structure, fully demonstrates the advantages of DRL[31-33], and effectively solves the problems in Compensation and Parameter adjustment control strategies.
  \item Based on the DRL control method for a single rectifier, a method using a single agent to simultaneously control two rectifiers is proposed. And this paper sets up the scenarios of running on flat ground, passing uphill/downhill sections, and passing through neutral sections for the next verification.
  \item To enable the rectifier to adapt to different operating conditions of EMUs, this paper proposes a reward shapingbased DRL control method by leveraging a reactive current command generator, and it is combined with the prioritized experience replay, which can improve the generalization ability of the control.
\end{enumerate}

\section*{II. DESIGN OF DRL CONTROLLER}
At present, the $d q$ decoupling control used in the CRH5 EMUs rectifier is to decouple the collected voltage and current on the input side of the rectifier through the second-order generalized integrator-phase locked loop (SOGI-PLL) to obtain their respective $d q$ components, and then control them according to different performance indexes of the $d q$ axis. This control strategy can be effectively applied to multiple-input multiple-output (MIMO) control systems [2], and the decoupled $d q$ current is approximately direct flow, which can realize PI control without steady-state error. The specific structure is shown in Fig. 1. [5].

In the traditional $d q$ decoupling control, the $d$-axis adopts a double closed-loop control structure, and the control goal is to stabilize the DC voltage $U_{\mathrm{dc}}$ at the reference value $U_{\mathrm{dc}}{ }^{*}$, which is set to 3600 V in CRH5. The $q$-axis adopts ordinary PI control, and the control goal is to make the q -axis current $i_{\mathrm{n} q}$ on the input\\
side of the rectifier approach the reference value $i_{\mathrm{n} q}{ }^{*}$, and $i_{\mathrm{n} q}{ }^{*}$ is generally set to 0A to achieve the unity power factor operation of the rectifier circuit. The specific current loop control block diagram is shown in Fig. 2 [5].

\begin{figure}[H]
\begin{center}
  \includegraphics[alt={},max width=\columnwidth]{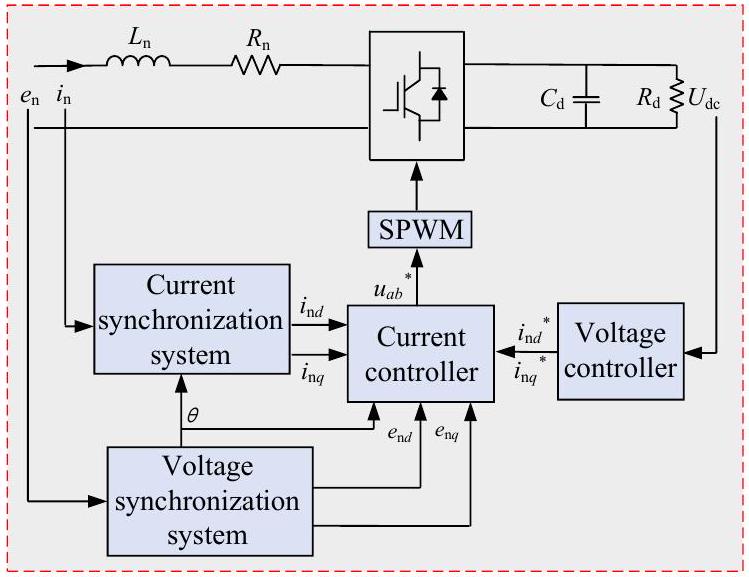}
\caption{Rectification structure of CRH5 EMUs based on $\boldsymbol{d} \boldsymbol{q}$ decoupling control.}
\end{center}
\end{figure}

\begin{figure}[H]
\begin{center}
  \includegraphics[alt={},max width=\columnwidth]{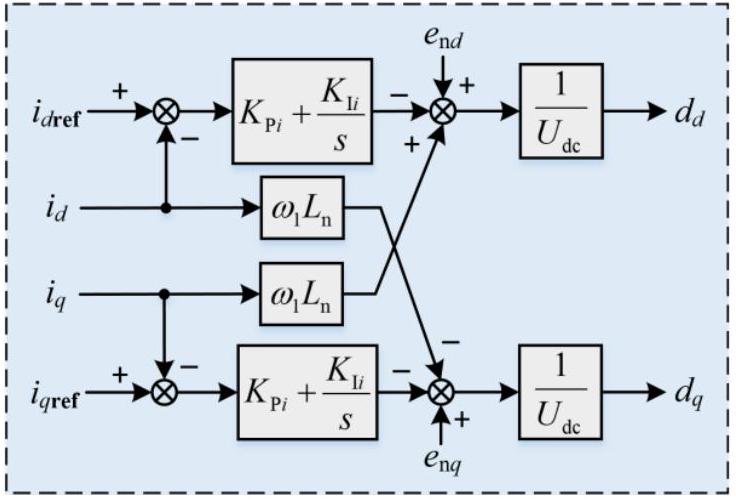}
\caption{CRH5 single-phase rectifier current loop control block diagram.}
\end{center}
\end{figure}

\section*{A. DRL replacement policy}
To fully demonstrate the performance advantages of DRL data-driven control and eliminate the shortcomings of PI control, it is considered to replace all PI controllers in the $d q$ decoupling control structure with agents, thereby eliminating the voltage loop and achieving parallel control of the $d q$ axis by agents [13].

A multi-input and dual-output neural network can be built inside the agent, and the two outputs of the network are taken as the respective control quantities of the $d q$ axis to realize the parallel control of the $d q$ axis.

Set the corresponding observations according to the control target of the $d q$ axis. According to the $d$-axis control target, the state of the $d$-axis can be set to the error between the reference value of $U_{\mathrm{dc}}$ and the actual value $U_{\mathrm{dc}}{ }^{*}-U_{\mathrm{dc}}$ (denoted as $U_{\mathrm{dc}}$ err) and its integral. According to the control target of the $q$-axis, the state of the $q$-axis can be set as the error $i_{\mathrm{n} q}{ }^{*}-i_{\mathrm{n} q}$ (denoted as $i_{\text{n } q}$ err) between the reference value and the actual value of the $i_{\text{n } q}$ and its integral value, where the function of the state integral is to correct the long-term stability of the system according to the accumulated value of the error. However, to ensure that the DRL agent can effectively recognize observations and avoid\\[0pt]
the gradient dispersion in the training process that will affect the convergence speed, it is also necessary to normalize the Score method [14]. The normalized data together form a multidimensional array, which can be used as the observation of the DRL agent.

According to the action space size and exploration efficiency of DRL, the two actions of the agent can be set as

\[
\left\{\begin{array}{c}
e_{\mathrm{n} d}-U_{d \mathrm{RL}}=u_{\mathrm{ab} d}^{*}  \tag{1}\\
U_{q \mathrm{RL}}=u_{\mathrm{ab} q}^{*}
\end{array}\right.
\]

where $U_{d \mathrm{RL}}$ represents one of the outputs of the agent, which can be used as the compensation voltage of $e_{\mathrm{n}} d$; $U_{q \mathrm{RL}}$ represents another one of the outputs of the agent, which can be used as the reference value of the $q$-axis voltage on the input side of the rectifier; $U_{d \mathrm{RL}}$ and $U_{q \mathrm{RL}}$ are both converted values. Other physical quantities are shown in Fig. 1.

Depending on the control target, the reward function can be set as

\begin{equation*}
r_{t}=-\left(Q_{1} \times U_{\mathrm{dc}} \mathrm{err}^{2}+Q_{2} \times i_{\mathrm{n} q} \mathrm{err}^{2}\right) \tag{2}
\end{equation*}

where $r_{t}$ represents single-step reward, and each error square term adopts the minimization of negative return based on the least square method to find the optimal strategy [18], to reduce the error between the actual value and the reference value. $Q_{1}$ and $Q_{2}$ represent the weight coefficients of each error square term in the single-step reward, and the control targets of both $d q$ axes need to be considered at the same time, so the sum of each error term should be added together.

To ensure the control accuracy of the rectifier circuit, the action space of the DRL should be continuous. Therefore, the Twin Delayed Deep Deterministic Policy Gradient (TD3) algorithm can be adopted, which is an improved version of the Deep Deterministic Policy Gradient (DDPG) algorithm. DDPG is an extended version of Deep Q-Network, which learns both policies and Q-value at the same time. By adding a policy network based on Q-Network, deterministic policies can be made, thus expanding into continuous action space. However, the Q-function that has been learned by DDPG usually overestimates the Q-value, destroying the policy [15]. TD3 can solve the above problems of the DDPG by introducing clipped double Q-Learning, delayed policy updates, and target policy smoothing [15]. Fig. 3 shows how the DRL actions, observations, and reward function constitute the DRL controller.

\begin{figure}[H]
\begin{center}
  \includegraphics[alt={},max width=\columnwidth]{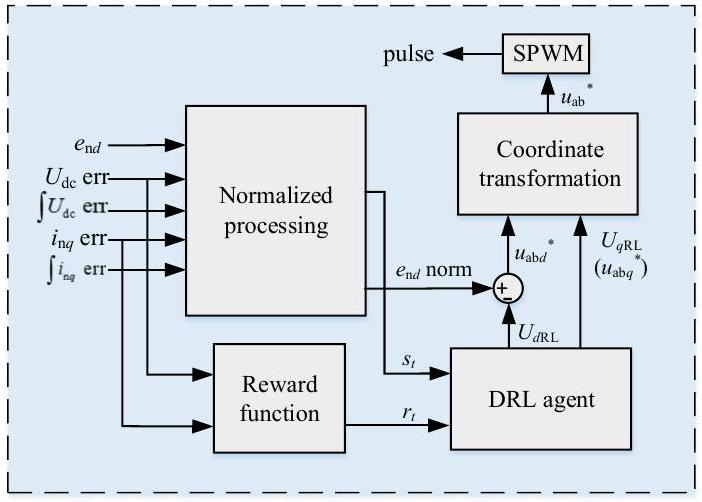}
\caption{Internal structure block diagram of DRL controller.}
\end{center}
\end{figure}

states, which can be adopted by using the Min-Max Normalization method or z-

The DRL controller is applied to the rectification structure of CRH5 EMUs, as shown in Fig. 4. In the figure, the switching frequency of the IGBT is 250 Hz as specified by the national standard, which is already a very low frequency. Therefore, the proposed DRL control can be applied to higher switching frequencies.

\begin{figure}[H]
\begin{center}
  \includegraphics[alt={},max width=\columnwidth]{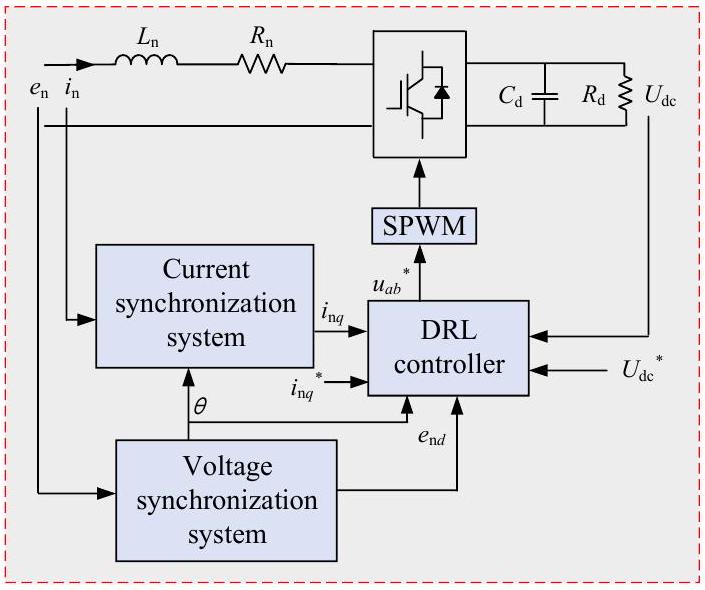}
\caption{Rectification structure of CRH5 EMUs based on DRL control.}
\end{center}
\end{figure}

As shown in Fig. 5, the specific network architectures of the Actor and Critic used in this paper are presented as follows.

\begin{figure}[H]
\begin{center}
  \includegraphics[alt={},max width=\columnwidth]{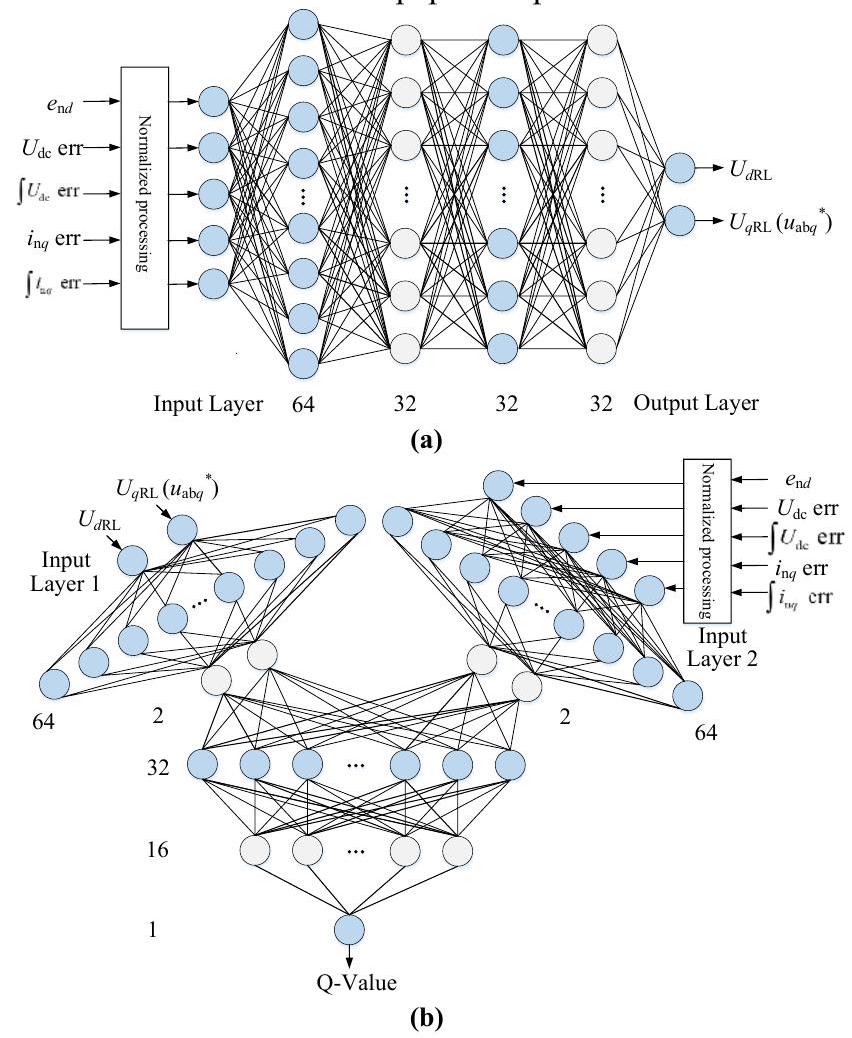}
\caption{(a) Actor network architecture; (b) Critic network architecture.}
\end{center}
\end{figure}

The training protocol defines the reinforcement learning process, including environment setup and workflow: Uses the Simulink environment, with 8 -dimensional observations and 2 -dimensional actions. Random seed is fixed via rng(0) to ensure consistent results. The Actor employs a deterministic policy with tanh output for action scaling. The Critic uses a twin Q-network structure to mitigate overestimation. Episode-based training uses experience replay (sampling 512 samples per update) and soft target network updates (every 10 steps with $\tau=0.005$). Noise-driven exploration with decaying variance, combined with target policy smoothing, improves generalization. Early stopping based on reward threshold (0) or maximum episodes (1000). Key hyperparameters and their functions are shown in TABLE I, as follows:

\begin{table*}[!t]
\centering
\caption{Comparison of DC Voltage Performance Indexes under Stable Working Conditions}
\resizebox{\textwidth}{!}{%
\begin{tabular}{|l|l|l|}
\hline
Actor learning rate 0.001 & Critic learning rate 0.0001 & gradient clipping threshold for the Actor 1 \\
\hline
L2 regularization factor 0.001 & \begin{tabular}{l}
discount factor \\
0.995 \\
\end{tabular} & \begin{tabular}{l}
experience replay buffer size \\
$1 \times 106$ \\
\end{tabular} \\
\hline
\begin{tabular}{l}
mini-batch size \\
512 \\
\end{tabular} & \begin{tabular}{l}
target update frequency \\
10 steps \\
\end{tabular} & \begin{tabular}{l}
target network smooth factor $(\tau)$ \\
0.005 \\
\end{tabular} \\
\hline
\begin{tabular}{l}
Initial exploration variance 1 \\
0.05 \\
\end{tabular} & \begin{tabular}{l}
decay rate 1 \\
$2 \times 10-4$ \\
\end{tabular} & \begin{tabular}{l}
minimum of variance 1 \\
0.001 \\
\end{tabular} \\
\hline
\begin{tabular}{l}
target policy smoothing initial variance 2 \\
0.1 \\
\end{tabular} & \begin{tabular}{l}
decay rate 2 \\
0.0001 \\
\end{tabular} &  \\
\hline
\end{tabular}
}
\end{table*}

\section*{B. Rectifier control scenario application}
Firstly, the control of the dual rectifiers is considered. Due to the limited power transmission of traction transformers, to increase the input power, the rectifier in the traction transmission system of actual EMUs is usually double [2], as shown in Fig. 6.

\begin{figure}[H]
\begin{center}
  \includegraphics[alt={},max width=\columnwidth]{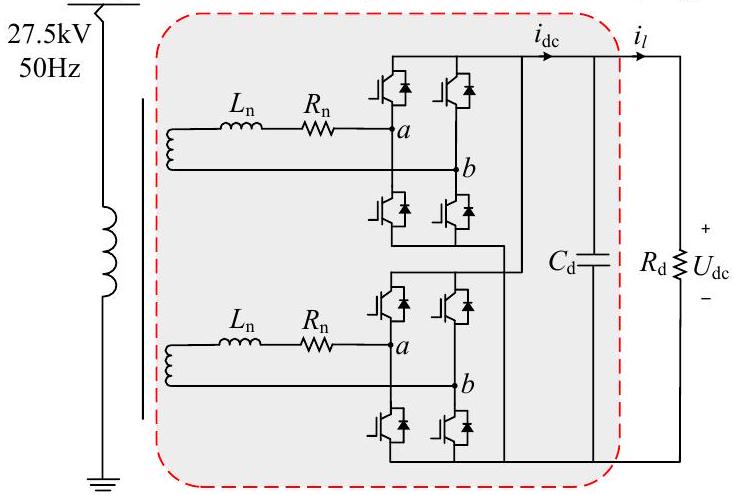}
\caption{Topology structure of dual rectifiers.}
\end{center}
\end{figure}

The traditional control method is that two rectifiers have independent control systems to control them separately, and the carrier phase-shifted PWM technology is used to reduce the high harmonic content [16]. To avoid problems such as poor training effect caused by excessive agent action and double-agent game, a method of controlling two rectifiers with a single agent is proposed in this paper.

In the traditional $d q$ decoupling control, the error between the actual value and the reference value of the DC voltage of $e_{\text{nd }}$ through the double closed-loop control structure. If the $d$ axis of two rectifier control systems adopts a single closedloop control structure, as long as the PI parameters correspond to the same, the compensated voltage of $e_{\mathrm{n} d}$ which are output by the two control systems will be the same.

\begin{figure}[H]
\begin{center}
  \includegraphics[alt={},max width=\columnwidth]{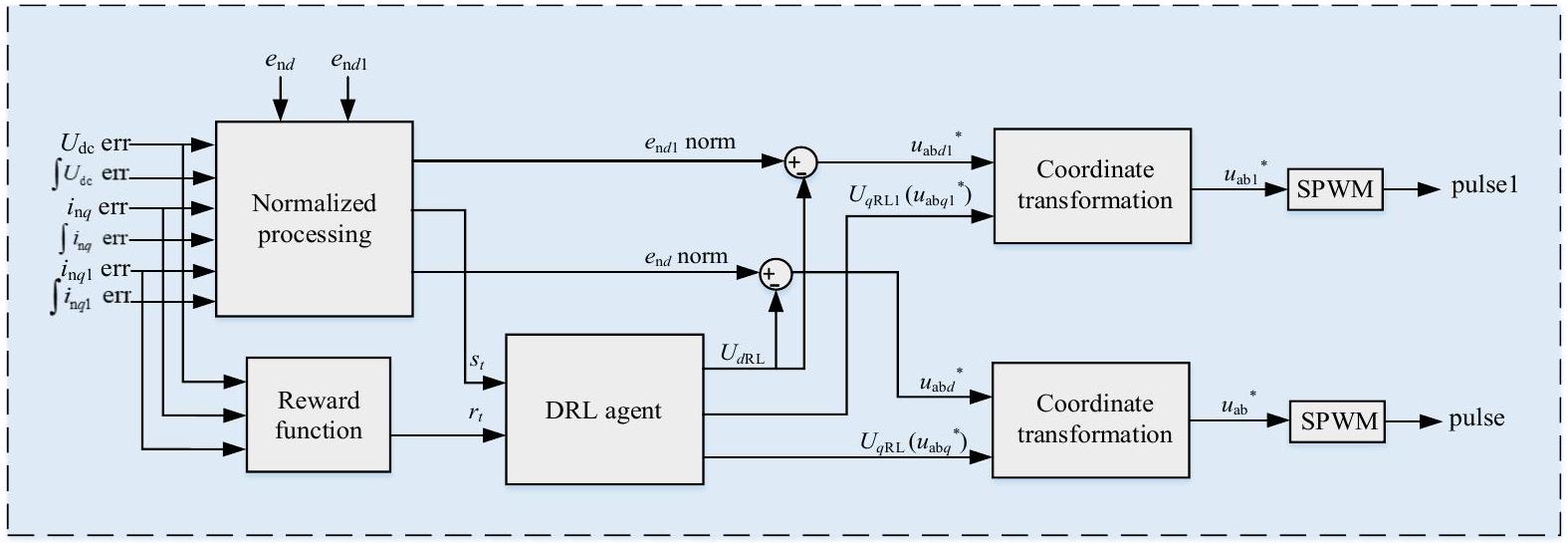}
\caption{DRL controller internal structure block diagram of dual rectifiers}
\end{center}
\end{figure}

Similarly, the $d$-axis in the DRL controller can be regarded as a single closed-loop control structure, then the $e_{\mathrm{n} d}$ of the two rectifier circuits can share one compensation voltage, according to (1), in different cases of $e_{\mathrm{n} d}$, different $u_{\mathrm{abd}}{ }^{*}$ can be obtained. In addition, to reduce fluctuations in the input current, DC voltage, and load current of the two rectifiers of the traction transformer, it is advisable to avoid the output current of the two rectifiers being the same as much as possible. In addition to the carrier phase-shifted PWM technology, the two $q$-axis currents (recorded as $i_{\mathrm{n} q}, i_{\mathrm{n} q 1}$ ) should be separately controlled. Therefore, the number of DRL actions can be set to 3, and two sets of $q$-axis observations can be set, as shown in Fig. 7. The error term corresponding to $i_{\mathrm{n} q 1}$ should also be\\
added to the reward function.\\
Secondly, the rectifier control of the multi-condition operation is considered. To ensure the stability of DC voltage in the whole operation process of EMUs, it is necessary to control the rectifier under different working conditions. This paper mainly analyzes the rectifier control of EMUs in four typical working conditions, namely running on flat ground, uphill, downhill, and passing a neutral section. The running speed of $200 \mathrm{~km} / \mathrm{h}$ is maintained during switching, and the absolute value of the slope parameter is $20 \%$.

When the EMUs are going up and down the hill, it will introduce the impact of the forward and reverse power flow caused by traction or braking, resulting in the fluctuation of\\[0pt]
traction supply catenary voltage and catenary current, thus causing the DC voltage fluctuation [17], as shown in Fig. 8.

\begin{figure}[H]
\begin{center}
  \includegraphics[alt={},max width=\columnwidth]{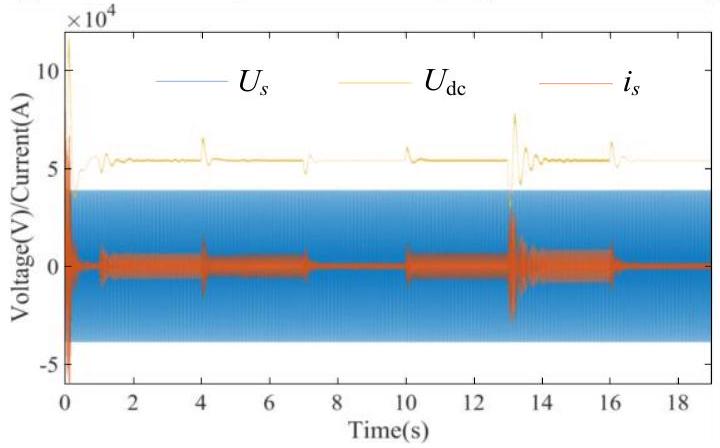}
\caption{Waveform of voltage current and DC voltage on the grid side when the EMUs go up and down the slope.}
\end{center}
\end{figure}

At this time, $i_{\mathrm{n} q}{ }^{*}$ can be set as a negative or positive value, then the power factor angle $\varphi$ on the input side of the rectifier is within the range of $0^{\circ} \sim 90^{\circ}$ or $-90^{\circ} \sim 0^{\circ}$. Thus, the current in the input side of the rectifier leads or lags the voltage $e_{\mathrm{n}}$, and the pulse rectifier can be equivalent to a capacitive or inductive load, thereby increasing or reducing the traction power supply catenary voltage to restore to the rated value [18]. At this time, the DC voltage $U_{\mathrm{dc}}$ should still be kept stable. However, the simulation test found that when the absolute value of the slope parameter is $20 \%$, the traction power supply catenary voltage only decreases or increases by about 80 V , and the catenary voltage is always in the range of $22.5-29 \mathrm{kV}$, and the EMUs can exert its rated power [19]. Therefore, the catenary voltage fluctuation can be ignored, and $i_{\mathrm{n} q}{ }^{*}$ is still set at 0 A and the operation is close to the unity power factor. However, from the harmonic propagation mechanism of the traction transmission system [20], it can be seen that the voltage of the traction network and the fluctuation of DC voltage are mutually influenced. Therefore, to maintain the stability of the catenary voltage and the DC voltage as much as possible, the control performance of the $q$-axis should not be over-considered when designing the DRL-based control system, but the stability of the DC voltage should be fully guaranteed.

The neutral section of the traction network is usually set on the level, but the slope is more than the level in the mountains, and the neutral section is mostly set on the slope. Therefore, the EMUs passing over the neutral section can be divided into three categories: passing the neutral section on flat ground, passing the neutral section uphill, and passing the neutral section downhill. After the pantograph breaker is switched off, the EMUs are in the no-power operation. The DC voltage should be regulated by the characteristics of the capacitor to maintain its stability [21]. Then the motor loses power and the rotor continues to rotate by inertia. All pulses should also be blocked before opening the breaker, then the control strategy will fail briefly [21], so it is unnecessary to consider the converter control at this stage.

After the pantograph circuit breaker is re-closed, the EMUs are re-powered, the phase is locked, the pulse blockade is lifted, and the capacitor is put into use again. At this time, the EMUs can be regarded as restarting, the difference is that the capacitor has the initial voltage and the motor has the initial\\
speed. The re-input of the phase-locked loop and the control loop will cause a large fluctuation in the DC voltage. In addition, if the EMUs end up passing the neutral section uphill/downhill and restart, the traction transmission system will also bear the impact of the forward or reverse power flow, which will further aggravate the fluctuation of the DC voltage. For example, under the PI control, there will be a large overshoot, as shown in Fig. 9.

\begin{figure}[H]
\begin{center}
  \includegraphics[alt={},max width=\columnwidth]{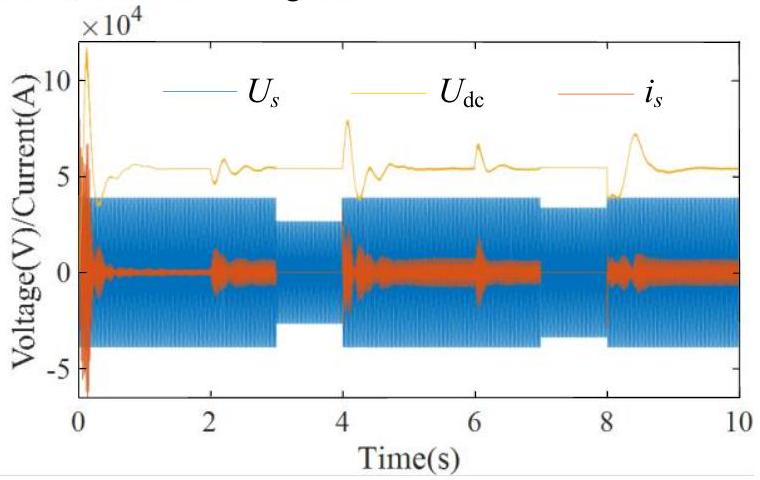}
\caption{DC voltage and waveform of voltage current at the side of the catenary when EMUs pass the neutral section.}
\end{center}
\end{figure}

The simulation test shows that the DRL controller can reduce the overshoot of DC voltage during the start of EMUs, and the stabilization time is short. Therefore, it is also suitable for the rectifier control of EMUs when passing the neutral section. The detailed control strategy is described in the following section.

The actual EMUs' operation process may include various operating conditions, therefore, DRL controllers need to have high generalization ability to adapt to different working conditions and their switching. At present, the typical processing method is presented below. First, different policy networks are trained under different working conditions, and then the Meta-Reinforcement Learning or Ensemble Learning algorithm is used to "fuse" these policy networks. However, meta-reinforcement learning first needs to find the common optimal parameters in many related tasks[22], which is not suitable for tasks with huge parameter differences. Ensemble Learning needs to combine multiple agents to train model weights, the system is large, and both algorithms need pretraining, which is a complicated process.

Deep Reinforcement Learning is based on Markov chains and is used to solve the sequential decision problem, that is, the state and action of the previous moment will affect the state and action of the next moment [22]. Therefore, the policy network in the agent can be considered to be determined by the "state chain" in the entire training process. Therefore, various working conditions can be added to the training of a single episode, and the agent can be trained to make continuous decisions in the whole process. However, this method has the problem of the sequence of different working conditions. To solve this problem, this paper proposes a "one-episode allsituation training method", as shown in Fig. 10.

In Fig. 10, considering that the EMUs keep a constant speed, the altitude of EMUs when passing uphill/downhill sections also changes uniformly. In the simulation test, the

\begin{figure}[H]
\begin{center}
  \includegraphics[alt={},max width=\columnwidth]{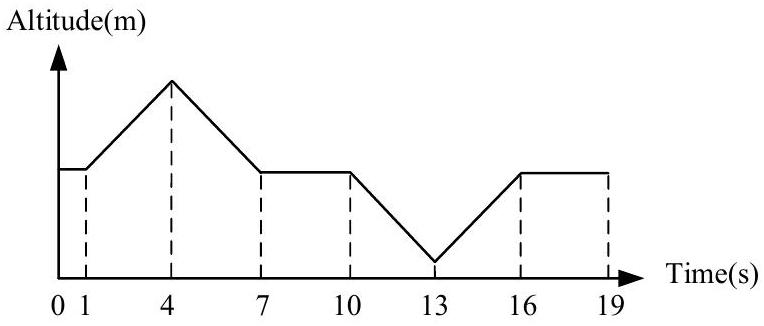}
\caption{Operation process of "One-episode all-situation training method".}
\end{center}
\end{figure}

initial 1s is the start time of EMUs on flat ground, and the time of EMUs on the hill and flat ground is set to 3s. The neutral section of EMUs can be randomly set in the EMUs' uphill, downhill, and flat opreation process. The entire process includes all situations of "uphill to downhill", "downhill to flat", "flat to downhill", "downhill to uphill", and "uphill to flat", to solve the sequence of different working conditions. Since the agent has been trained in all cases, it can be applied to any working condition and switching, with high generalization ability.

\section*{III. Improvement and Verification of DRL Controller}
In control systems, the bandwidth reflects the system's tracking capability and response speed to input signals. Therefore, the control performance and stability of a PI closedloop system are related to its bandwidth. In practical applications, the bandwidth of a PI controller is determined by its selected parameters. Traditional dq current decoupling control employs a dual closed-loop control structure, where the selection and coordination of PI parameters in the voltage loop and current loop determine the bandwidth of each loop and the overall system, thereby affecting control performance and stability. The DRL control method proposed in this paper eliminates the voltage loop and replaces all PI controllers in the current loop, achieving favorable control performance as shown in Fig. 10. This approach can be approximately regarded as having found globally optimal parameters, thus obviating the need for manual PI parameter tuning and resolving the bandwidth limitation issue inherent in traditional PI control.

The simulation results show that the DRL-based dual rectifiers have good control performance under a single stable working condition, as shown in Fig. 11. Compared with PI control, the performance indicators of the DC voltage are shown in TABLE II.

\begin{figure}[H]
\begin{center}
  \includegraphics[alt={},max width=\columnwidth]{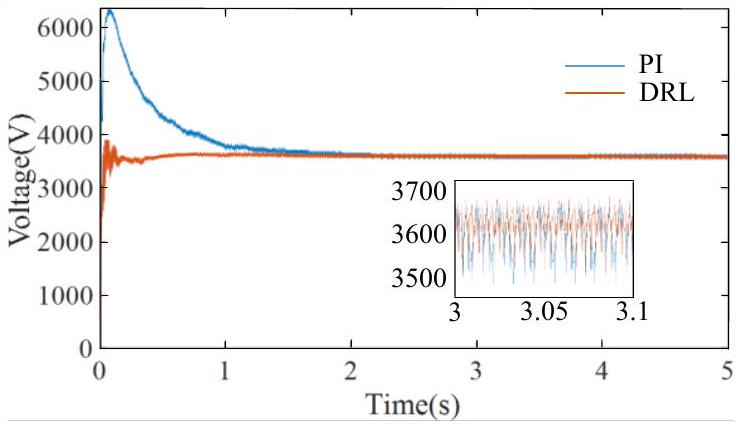}
\caption{Comparison of DC voltage waveforms under stable conditions.\\
TABLE II}
\end{center}
\end{figure}

\begin{table}[H]
\begin{center}
\caption{Comparison of DC Voltage Performance Indexes under Stable Working Conditions}
\begin{tabular}{ccccc}
\hline
Control & \begin{tabular}{c}
Overshoot \\
(\%) \\
\end{tabular} & THD & \begin{tabular}{c}
Adjustment \\
time(s) \\
\end{tabular} & \begin{tabular}{c}
Voltage \\
Fluctuation(V) \\
\end{tabular} \\
\hline
PI & $84.72 \%$ & $0.73 \%$ & 1.8 s & $\pm 80 \mathrm{~V}$ \\
DRL & $6.94 \%$ & $0.45 \%$ & 0.7 s & $\pm 70 \mathrm{~V}$ \\
\hline
\end{tabular}
\end{center}
\end{table}

At the same time, this method can also achieve good control performance of $i_{\mathrm{n} q}$, as shown in Fig. 12.

\begin{figure}[H]
\begin{center}
  \includegraphics[alt={},max width=\columnwidth]{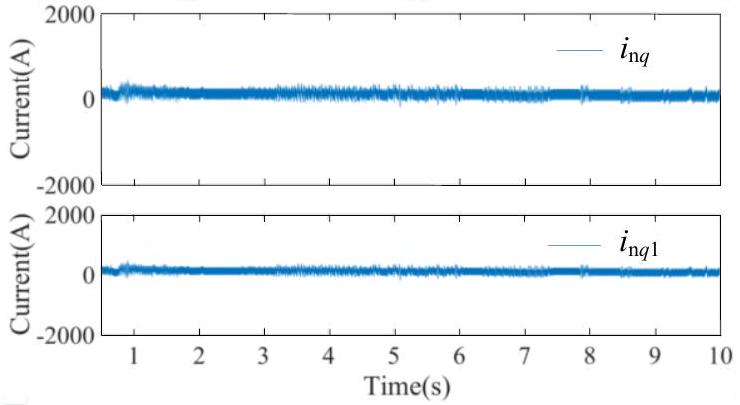}
\caption{Current waveforms of two Q-axes under stable conditions.}
\end{center}
\end{figure}

The training finally converged at the 155st episode. In Simulink, each episode is set to 10 seconds with a step size of 0.001 seconds, containing 10,000 steps and taking 5 minutes. Therefore, the total duration is 15 hours and 54 minutes. The converged Episode Reward is -160 .

However, the simulation results show that the control performance is poor if the original control strategy based on DRL is used in various operating conditions. The reason is that actions of DRL cannot achieve large jumps during the process of switching operating conditions.

\section*{A. Reward Shaping}
Due to the use of a "unified reward function": the difference in single-step reward $r_{t}$ under different conditions and switching (i.e. different states) is very small, the limited exploration is caused. Therefore, each step tends to be the same and cannot be significantly changed. This problem can be solved by using Reward Shaping [24], by establishing a "nonunified reward function". The reward function in the DRL controller of the dual rectifiers can be set to

\begin{equation*}
r_{t}=-\left(Q_{1} \times U_{\mathrm{dc}} \mathrm{err}^{2}+Q_{2} \times i_{\mathrm{nq}} \mathrm{err}^{2}+Q_{3} \times i_{\mathrm{nq1}} \mathrm{err}^{2}\right) \tag{3}
\end{equation*}

where $Q_{1}, Q_{2}$, and $Q_{3}$ are the corresponding weight coefficients of each error term, respectively. The two control objectives of the dq-axis contradict each other in Fig. 13.

\begin{figure}[H]
\begin{center}
  \includegraphics[alt={},max width=\columnwidth]{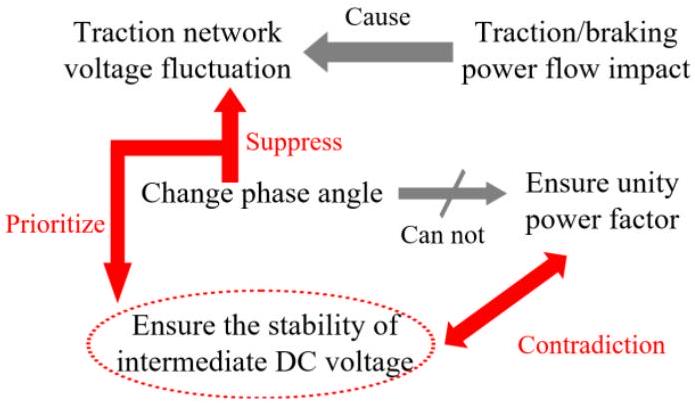}
\caption{The contradiction of control objectives in dq axes.}
\end{center}
\end{figure}

It can also be seen in Section II.B that when the EMUs switch between flat ground running, uphill, and downhill, it is not\\
necessary to pay too much attention to whether $i_{\mathrm{n} q}$ is 0 A , but more attention should be paid to the stability of DC voltage $U_{\mathrm{dc}}$. Therefore, when the working condition is switched, the reward function can appropriately increase $Q_{1}$ and decrease $Q_{2}$ and $Q_{3}$. To reflect the coupling relationship between each control performance, $Q_{1}+Q_{2}+Q_{3}$ can also be set to $Z$, where $Z$ is a fixed value, thus they are restricted to each other.

Accordingly, when $U_{\mathrm{dc}}$ fluctuates, $U_{\mathrm{dc}}$ err increases, and $Q_{1}$ increases. Therefore, the weight coefficients of each error item in $r_{t}$ can be assigned according to the size of $U_{\mathrm{dc}}$ err, as shown in Fig. 14.

With the increase of $U_{\mathrm{dc}}$ err, the corresponding weight coefficient $Q_{1}$ also increases. And the smaller the $U_{\mathrm{dc}}$ err, the faster $Q_{1}$ declines, which is conducive to increasing $r_{t}$ and better making $U_{\mathrm{dc}}$ err approach 0 . According to Fig. 14, a logarithmic function is used to fit $Q_{1}$.

\begin{figure}[H]
\begin{center}
  \includegraphics[alt={},max width=\columnwidth]{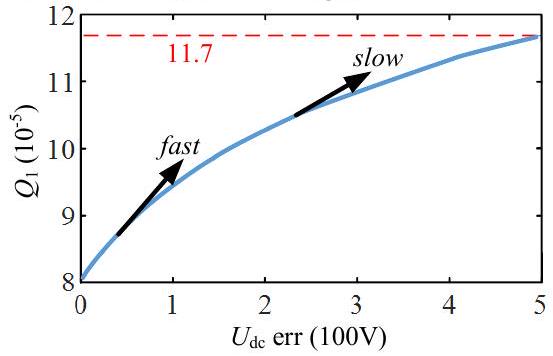}
\caption{Q1 of Reward Shaping.}
\end{center}
\end{figure}

\begin{equation*}
Q_{1}=\log _{1.6229}(x+1)+8 \tag{4}
\end{equation*}

And $i_{\mathrm{n} q}$ and $i_{\mathrm{n} q 1}$ are at the same level, $Q_{2}=Q_{3}, Z=12$, so

\begin{equation*}
Q_{2}=Q_{3}=\frac{1}{2}\left(Z-Q_{1}\right)=\frac{1}{2}\left(4-\log _{1.6229}(x+1)\right) \tag{5}
\end{equation*}

Changing the weight coefficient is equivalent to introducing additional Reward functions based on potential into each error term [24]. After completing Reward Shaping, 50 episodes of training are carried out on the agent, and it is found that the action can jump significantly during the switching of each working condition, as shown in Fig. 15. This shows that the agent can adapt to recognize the working condition switching.

\begin{figure}[H]
\begin{center}
  \includegraphics[alt={},max width=\columnwidth]{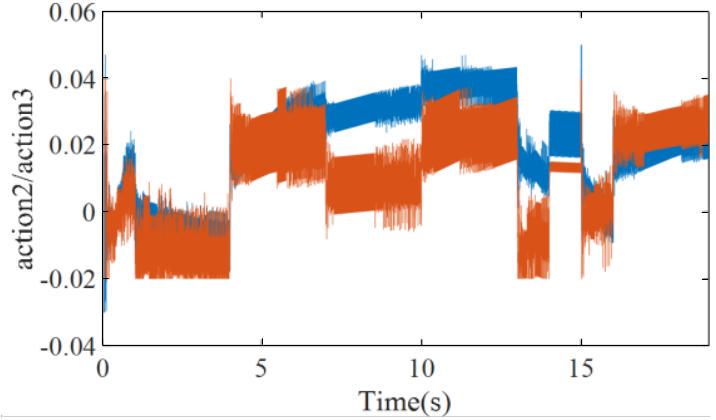}
\caption{Agent's 2-way actions after 50 episodes of training.}
\end{center}
\end{figure}

\section*{B. Prioritized Experience Replay}
The experiment shows that although Reward Shaping can improve the control performance based on DRL to a certain extent. Due to too many working conditions and switches set in a single training episode (see Fig. 10), it is difficult for the agent to distinguish states between many working conditions\\
and switches, resulting in large fluctuations in the DC voltage in the later period, as shown in Fig. 16.

\begin{figure}[H]
\begin{center}
  \includegraphics[alt={},max width=\columnwidth]{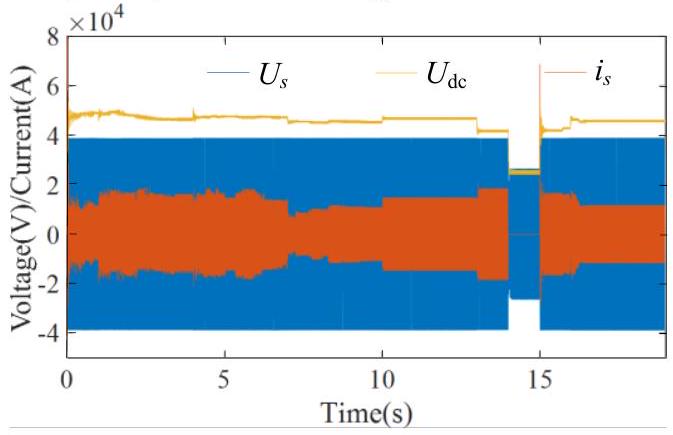}
\caption{Waveform of voltage and current at the catenary side and intermediate DC voltage after Reward Shaping.}
\end{center}
\end{figure}

The reason is that a large amount of experience is stored in the experience buffer, and the TD3 algorithm distributes samples evenly in the experience. It is easy to extract inappropriate experiences during training to update each network, resulting in a delay in updating the expected network. Further, the convergence of episode reward is poor, and the number of episodes is also large.

Prioritized experience replay (PER) is to add the TD error of the current state and the priority $P$ of the current experience to the original experience, which can be used to solve this problem. The smaller the TD error, the greater the value of the experience, and the greater the $P$ should be, so the priority $P$ of the experience is calculated as follows [25].

\begin{equation*}
P(i)=\frac{p_{i}^{\alpha}}{\sum_{k} p_{k}^{\alpha}}, \quad p_{i}=\left|\delta_{i}\right|+\varepsilon \tag{6}
\end{equation*}

where $p_{i}$ refers to the priority of state $i, \delta_{i}$ refers to the TD error of state $i$, and $\varepsilon$ is a small value to prevent the experience of $\delta_{i} =0$ from being extracted with a probability of 0 , thus avoiding network overfitting.

Prioritized experience replay can solve the inefficiency problem caused by distributing samples evenly in the experience buffer of the TD3 algorithm, greatly reducing the number of network updates, and avoiding the sudden decrease of episode reward in the training process, shortening the convergence time. In addition, similar to Deep Learning, data preprocessing is required before training, such as regularization and smoothing, while prioritized experience replay can be equivalent to regularization and smoothing by prioritizing experiences in the experience buffer.

Reward shaping is based on potential function theory. By designing auxiliary rewards (such as state potential differences), it ensures consistency with the original target strategy and meets the "admissibility" condition to avoid policy distortion. Additionally, it employs dynamic weights to guide the agent to prioritize critical states, enhancing the directionality and efficiency of the agent's learning. Prioritized experience replay relies on importance sampling to correct biases, uses weight decay coefficients to ensure

\begin{figure}[H]
\begin{center}
  \includegraphics[alt={},max width=\columnwidth]{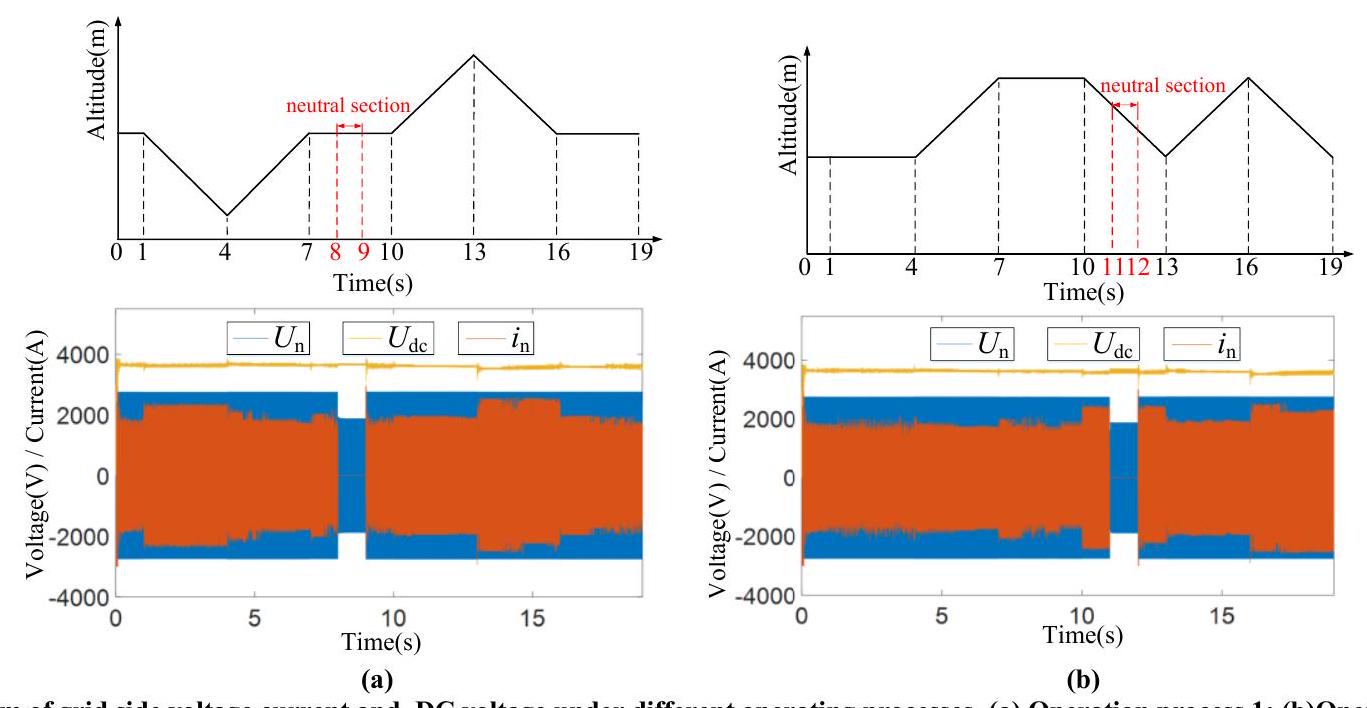}
\caption{Waveform of grid side voltage current and DC voltage under different operating processes. (a) Operation process 1; (b)Operation process 2.}
\end{center}
\end{figure}

unbiased updates, and combines the coverage theorem to prevent critical experiences from being overlooked. Its stability aligns with TD convergence in tabular MDPs, while function approximation scenarios require network regularization and target network mechanisms. Together, they guarantee the effectiveness of DRL model outputs through theoretical constraints such as MDP policy equivalence and sample coverage.

The other two operating processes different from Fig. 10. can be designed to verify the generalization ability of the improved DRL controller. The control effect of the DC voltage is shown in Fig. 17. It can be seen that the agent obtained based on the one-episode all-situation training method has a strong generalization ability and can be used for vehicle rectifier control under various working conditions and their switching.

As shown in Fig. 18, to further verify the training differences among three schemes - before adding reward shaping, after adding reward shaping, and after adding both reward shaping and prioritized experience replay - this paper presents the convergence curve of the episode reward. The corresponding final convergence time and single-episode reward under the training of the three schemes are shown in TABLE III. It can be seen that both reward shaping and prioritized experience replay can improve the final reward.

\begin{figure}[H]
\begin{center}
  \includegraphics[alt={},max width=\columnwidth]{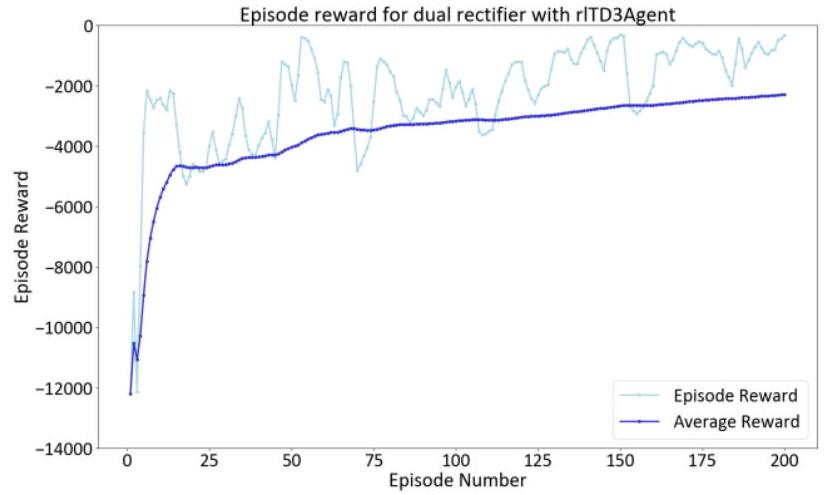}
\caption{(a)}
\end{center}
\end{figure}

\begin{figure}[H]
\begin{center}
  \includegraphics[alt={},max width=\columnwidth]{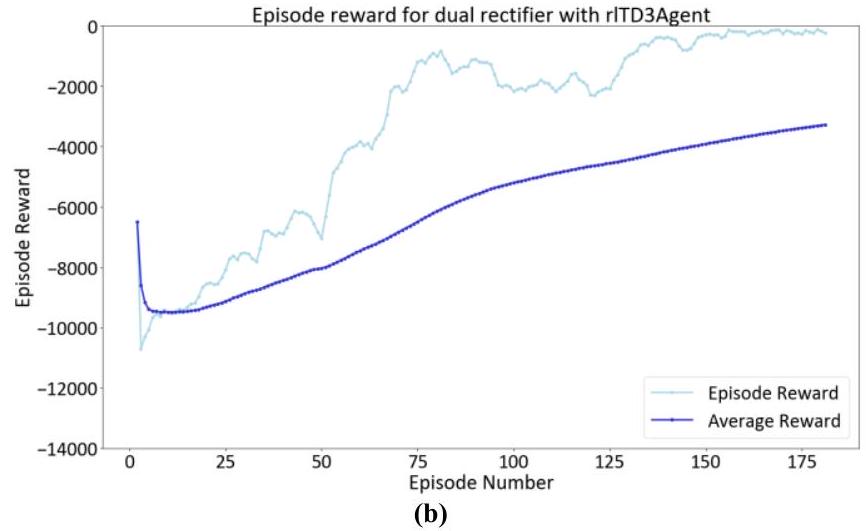}
\caption{Episode reward. (a) Without prioritized experience replay; (b) Adding prioritized experience replay}
\end{center}
\end{figure}

\begin{table}[H]
\begin{center}
\caption{The Final Convergence Time and Single-Episode Reward under Training with Different Schemes}
\begin{tabular}{|l|l|l|}
\hline
Scheme & Time & Reward \\
\hline
Without Reward Shaping & 28 hours and 57 minutes & -450 \\
\hline
With Reward Shaping & 25 hours and 24 minutes & - 310 \\
\hline
With Reward Shaping and Prioritized Experience Replay & 22 hours and 49 minutes & -190 \\
\hline
\end{tabular}
\end{center}
\end{table}

\section*{C. Comparison of other control effects}
Due to the nonlinear characteristics of the rectifiers in EMUs, and the feedback linearization control and model predictive control need to rely on precise mathematical models when linearizing the system, problems such as changes in reference trajectories or model mismatches may occur, leading to poor system performance. Therefore, nonlinear controls such as sliding mode control, robust control and passive control are introduced. However, since these nonlinear controls are all based on the energy perspective, there exist problems such as buffering and poor steady-state accuracy. The intelligent control of existing converters, such as fuzzy control, expert control, and genetic algorithms, do not rely on precise mathematical models and have good adaptive capabilities. However, the design is relatively complex, and the precise\\
analytical solution of the control output cannot be obtained. Both the control accuracy and the steady-state accuracy are poor.

As shown in Fig. 19, although the above-mentioned various controls perform well in the corresponding literature, but only simplify the load to a variable resistor, this paper only shows the poor control effects of model predictive control, sliding mode control and fuzzy control under extreme conditions such as the load being equivalent to a current source that can reflect the tidal current impact when passing uphill/downhill.

\begin{figure}[H]
\begin{center}
  \includegraphics[alt={},max width=\columnwidth]{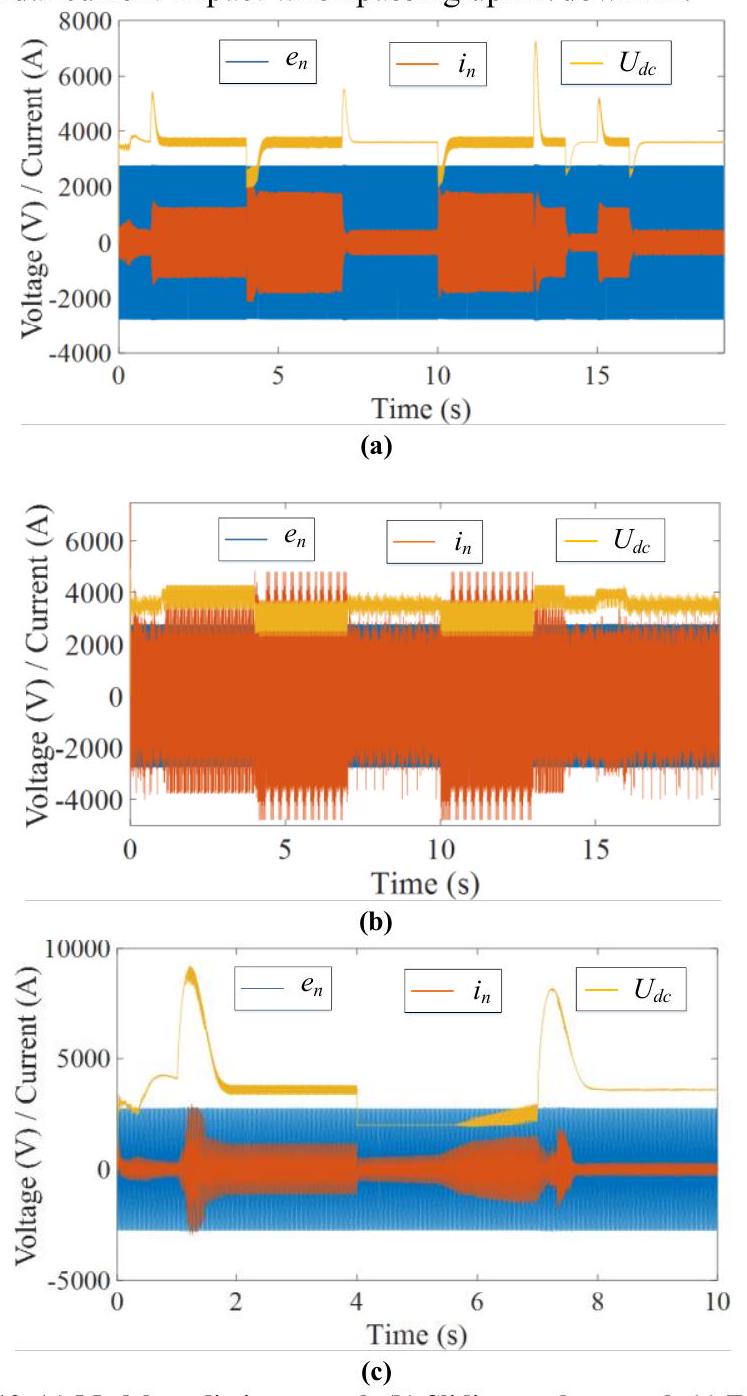}
\caption{(a) Model predictive control; (b) Sliding mode control; (c) Fuzzy control}
\end{center}
\end{figure}

\begin{table}[H]
\begin{center}
\caption{Comparison of Computational Overhead for Various Control Methods}
\begin{tabular}{|l|l|l|l|}
\hline
Control Method & Voltage fluctuation & Second harmonic amplitude & THD \\
\hline
MPC & $2000 \mathrm{~V} \sim 7500 \mathrm{~V}$ & 200 V & 1.2\% \\
\hline
SMC & $2500 \mathrm{~V} \sim 4200 \mathrm{~V}$ & 1000 V & 7\% \\
\hline
Fuzzy Control & $2500 \mathrm{~V} \sim 8500 \mathrm{~V}$ & 140 V & 0.84\% \\
\hline
\end{tabular}
\end{center}
\end{table}

The performance indicators of the intermediate DC voltage for the three control methods in the above figure are shown in TABLE IV.

The computational overheads of other conventional control methods are shown in TABLE V.

\begin{table}[H]
\begin{center}
\caption{Comparison of Computational Overhead for Various Control Methods}
\begin{tabular}{|l|l|l|l|}
\hline
Control Method & Resource Consumption & Time Cost & Response Time \\
\hline
 & High computational resource requirements (requires & Long training time (possibly days to months) &  \\
\hline
DRL Control & \begin{tabular}{l}
GPU/TPU) \\
Large-scale data storage requirements High memory usage during training \\
\end{tabular} & Moderate inference speed (depends on model complexity) Frequent online updates required & 10\~{}100 ms \\
\hline
Model Predictive Control & Medium computational resources (depends on model complexity) Requires real-time state estimation Online optimization solver & Computational time proportional to prediction horizon High real-time requirements necessitate model simplification Optimization solution may be time-consuming & $5 \sim 50 \mathrm{~ms}$ \\
\hline
Sliding Mode Control & Low computational resource requirements (only requires state feedback) & Fast computation speed (only simple mathematical operations) & 0.1\~{}0.5 ms \\
\hline
 & No complex model storage needed Simple parameter adjustment & Strong real-time performance Negligible delay &  \\
\hline
Fuzzy Control & Low to medium computational resources (depends on rule base size) Rule base storage requirements Membership function calculations & Medium computation speed (rule inference requires time) Can be optimized via look-up tables More rules increase computation time & $1 \sim 10 \mathrm{~ms}$ \\
\hline
\end{tabular}
\end{center}
\end{table}

\section*{D. Stability analysis and robustness analysis}
Due to the complexity of the neural network, it is not easy to find the transfer function between its input and output. Therefore, Lyapunov's second method can be used for stability analysis to verify whether the control system can reach the equilibrium point from the perspective of energy. For the stability analysis of nonlinear systems, the Jacobian matrix method (also known as Krasovskii's method) can be used [26]. It can be obtained from (2) in [5].

\begin{equation*}
\left(0-i_{\mathrm{n} q}\right)=-\left(0-i_{\mathrm{n} d}\right) \cdot \omega+\frac{1}{L_{0}}\left(u_{\mathrm{ab} q}-u_{\mathrm{n} q}\right) \tag{7}
\end{equation*}

Combining (1) to convert the action to before conversion, then there is

\[
f_{\mathrm{ANN}}\left(\left[\begin{array}{c}
0-i_{\mathrm{n} q}  \tag{8}\\
3600-U_{\mathrm{dc}}
\end{array}\right]\right)=\frac{1}{3600}\left[\begin{array}{c}
u_{q \mathrm{RL}} \\
u_{d \mathrm{RL}}
\end{array}\right]
\]

where $f_{\text{ANN }}($.$) is a function between the input and output of the$ neural network. The control system does not contain $i_{\text{nd }}$ and can be replaced with $i_{\mathrm{n} d}=i_{\mathrm{n} q} \times \cot \varphi$ [18]. The $u_{\mathrm{n} q}$ obtained by SOGIPLL is 0 . Thus there is

\begin{align*}
& \left(0-i_{\mathrm{n} q}\right)=-\left(0-i_{\mathrm{n} q}\right) \cdot \cot \varphi \cdot \omega+\frac{3600}{L_{0}} \times  \tag{9}\\
& f_{\mathrm{ANN}}\left(\left[\begin{array}{c}
0-i_{\mathrm{n} q} \\
3600-U_{\mathrm{dc}}
\end{array}\right]\right)_{(1,1)}=f_{1}\left(\left[\begin{array}{c}
0-i_{\mathrm{n} q} \\
3600-U_{\mathrm{dc}}
\end{array}\right]\right)
\end{align*}

In the same way,

\[
\left(3600-U_{\mathrm{dc}}\right)=f_{2}\left(\left[\begin{array}{c}
0-i_{\mathrm{nq}}  \tag{10}\\
3600-U_{\mathrm{dc}}
\end{array}\right]\right)
\]

Take a single output of a fully connected neural network as an example, if $0-i_{\mathrm{n} q}=x_{1}, 3600-U_{\mathrm{dc}}=x_{2}$, then there is

\begin{equation*}
f_{\mathrm{ANN}}\left(x_{1}, x_{2}\right)_{(1,1)}=\underbrace{\sum \cdots \varphi}_{n-k-2} \varphi\left(\sum_{i=1}^{m_{1}} w_{p i}^{(k+1)} w_{p} \varphi\left(\sum_{j=1}^{m_{0}} w_{i j}^{(k)} x_{j}^{(k)}+b_{i}^{(k+1)}\right)+b_{p}^{(k+2)}\right) \tag{11}
\end{equation*}

where the neural network has a total of $n$ layers; $m_{0}$ represents a total of $m_{0}$ nodes in the $k$ layer; $x_{j}$ represents the input of the $k$ layer; $w_{i j}$ represents the weight between the $i$ node of the $k+1$ layer and the $j$ node of the $k$ layer; $b_{i}$ represents the bias of the $i$ node of the $k+1$ layer; $\varphi($.$) represents a nonlinear function$ after the node, where ReLu is non-differentiable at $x=0$, which can be LeakReLu or the tangent function; $m_{1}$ represents $m_{1}$ nodes in layer $k+1$; $w_{p i}$ represents the weight between the $p$ node in layer $k+2$ and the $i$ node in layer $k+1$; $b_{p}$ represents the bias of the $p$ node in layer $k+2$.

Let the control system function be

\[
\boldsymbol{f}_{\text{con }}=\left[\begin{array}{l}
f_{1}  \tag{12}\\
f_{2}
\end{array}\right]
\]

So $\boldsymbol{f}_{\text{con }}($.$) is differentiable from x_{i}(i=1,2, \ldots n)$. Therefore,

\[
\left[\begin{array}{c}
0  \tag{13}\\
0-i_{\mathrm{n} q} \\
3600-U_{\mathrm{dc}}
\end{array}\right]=\boldsymbol{f}_{\mathrm{con}}\left(\left[\begin{array}{c}
0-i_{\mathrm{n} q} \\
3600-U_{\mathrm{dc}}
\end{array}\right]\right)
\]

is differentiable from independent variables, too.\\
Relist the Jacobian matrix of the control system, as follows

\[
\boldsymbol{J}(x)=\frac{\partial \boldsymbol{f}_{\text{con }}(\boldsymbol{x})}{\partial \boldsymbol{x}}=\left[\begin{array}{ll}
\frac{\partial f_{1}}{\partial x_{1}} & \frac{\partial f_{1}}{\partial x_{2}}  \tag{14}\\
\frac{\partial f_{2}}{\partial x_{1}} & \frac{\partial f_{2}}{\partial x_{2}}
\end{array}\right]
\]

A sufficient condition for the asymptotically stable system at the origin is that given a positive definite real symmetric matrix $\boldsymbol{P}$, the following matrices are positive definite [26].

\begin{equation*}
\boldsymbol{Q}(\boldsymbol{x})=-\left\lfloor\boldsymbol{J}^{\mathrm{T}}(\boldsymbol{x}) \boldsymbol{P}+\boldsymbol{P} \boldsymbol{J}(\boldsymbol{x})\right\rfloor \tag{15}
\end{equation*}

Take $\boldsymbol{P}=\left[\begin{array}{cc}\lambda_{1} & 0 \\ 0 & \lambda_{2}\end{array}\right]$, and $\lambda_{1}, \lambda_{2}>0$, then

\[
\boldsymbol{Q}(\boldsymbol{x})=-\left[\begin{array}{cc}
2 \lambda_{1} \frac{\partial f_{1}}{\partial x_{1}} & \lambda_{1} \frac{\partial f_{1}}{\partial x_{2}}+\lambda_{2} \frac{\partial f_{2}}{\partial x_{1}}  \tag{16}\\
\lambda_{1} \frac{\partial f_{1}}{\partial x_{2}}+\lambda_{2} \frac{\partial f_{2}}{\partial x_{1}} & 2 \lambda_{2} \frac{\partial f_{2}}{\partial x_{2}}
\end{array}\right]
\]

If $k=1$, the layer $k$ is the input layer, then

\begin{align*}
& f_{\mathrm{ANN}}^{\prime}\left(x_{1}, x_{2}\right)_{(1,1)} \\
& =\underbrace{\varphi^{\prime} \cdots \sum^{\prime}}_{n-2} \varphi^{\prime}\left(\sum_{i=1}^{m_{1}} w_{p i} \varphi\left(\sum_{j=1}^{2} w_{i j} x_{j}+b_{i}\right)+b_{p}\right) .  \tag{17}\\
& \sum_{i=1}^{m_{1}} w_{p i} \varphi^{\prime}\left(\sum_{j=1}^{2} w_{i j} x_{j}+b_{i}\right) \cdot w_{i 1}
\end{align*}

Then, the leading principle minor of $\boldsymbol{Q}(\boldsymbol{x})$ is shown below.

\begin{equation*}
\resizebox{0.98\columnwidth}{!}{$\displaystyle
2 \lambda_{\mathbf{1}} \frac{\partial f_{1}}{\partial x_{1}}=2 \lambda_{\mathbf{1}}\left(\cot \varphi \cdot \omega-\frac{3600}{L_{\mathbf{0}}}\binom{\varphi^{\prime} \cdots \sum \sum_{n-2} \varphi^{\prime}\left(\sum_{i=1}^{m_{1}} w_{p i} \varphi\left(\sum_{j=1}^{2} w_{i j} x_{j}+b_{i}\right)+b_{p}\right)}{\cdot \sum_{i=1}^{m_{1}} w_{p i} \varphi^{\prime}\left(\sum_{j=1}^{2} w_{i j} x_{j}+b_{i}\right) \cdot w_{i 1}}\right)
$}\tag{18}
\end{equation*}

If it is greater than 0 , then

\begin{equation*}
\resizebox{0.98\columnwidth}{!}{$\displaystyle
\cot \varphi \cdot \omega-\frac{3600}{L_{0}}(\underbrace{\varphi^{\prime} \cdots \sum \varphi^{\prime}\left(\sum_{i=1}^{m_{1}} w_{p i} \varphi\left(\sum_{j=1}^{2} w_{i j} x_{j}+b_{i}\right)+b_{p}\right)}_{n-2} \underbrace{m_{1}}_{i=1} w_{p i} \varphi^{\prime}\left(\sum_{j=1}^{2} w_{i j} x_{j}+b_{i}\right) \cdot w_{i 1}{ })>0
$}\tag{19}
\end{equation*}

Take $\omega=100 \pi$, and $L_{0}=5.4 \times 10^{-3}$, then

\begin{equation*}
\resizebox{0.98\columnwidth}{!}{$\displaystyle
\binom{\underbrace{\varphi^{\prime} \cdots \sum}_{n-2} \varphi^{\prime}\left(\sum_{i=1}^{m_{1}} w_{p i} \varphi\left(\sum_{j=1}^{2} w_{i j} x_{j}+b_{i}\right)+b_{p}\right)}{\sum_{i=1}^{m_{1}} w_{p i} \varphi^{\prime}\left(\sum_{j=1}^{2} w_{i j} x_{j}+b_{i}\right) \cdot w_{i 1}}<4.712 \times 10^{-3} \cot \varphi
$}\tag{20}
\end{equation*}

If $\boldsymbol{Q}(\boldsymbol{x})$ of the $2 \times 2$ principle minor is also greater than 0 , then $\boldsymbol{Q}(\boldsymbol{x})$ is positive definite, and the control system is asymptotically stable at the origin.

When the EMUs travel at a constant speed on flat ground, it can be considered that the load is constant and unchanged. Then, the input impedance of the train is also constant and unchanged. There is no impact of forward and reverse power flow, the catenary voltage is stable, and the unity power factor operation can be achieved at this time, that is, $\varphi=0^{\circ}$. And $\cot 0^{\circ}= \pm \infty$, with the iteration of training, parameters such as weight, bias, input, and output of each layer of the neural network, can be changed, so that the gradient of the neural network can be positive or negative of finite size. Thus, (20) can be established. At this time, $\boldsymbol{Q}(\boldsymbol{x})$ is positive definite, and the control system is asymptotically stable at the origin. When the EMUs pass the neutral section or there is a brief external interference, the load and input impedance of trains will slightly change, and the unity power factor operation control is still in place. Under the influence of SOGI-PLL, although $i_{\mathrm{n} q}=0 \mathrm{~A}$ can still be controlled, (20) is still valid, $\boldsymbol{Q}(\boldsymbol{x})$ is still positive definite, and the control system is still asymptotically stable at the origin.

When the EMUs pass uphill/downhill sections, the load and the input impedance of trains will change, the impact of forward and reverse power flow will be introduced, and the catenary voltage fluctuation can be suppressed by changing $\varphi$. Then, the range of $\cot \varphi$ is $\left[0, R_{1}\right]$ on uphill and $\left[-R_{2}, 0\right]$ on downhill, where $R_{1}$ and $R_{2}$ are any finite positive real numbers. With the iteration of training, the weight, bias, and input and\\
output parameters of each layer of the neural network can be changed, so that the gradient of the neural network is just less than $4.712 \times 10^{-3} \cot \varphi$, and (20) still holds. However, the gradient value of the neural network cannot switch back and forth between the two value domains. As long as (20) is valid for downhill, it can be guaranteed for uphill, that is, the gradient is less than the negative value of $4.712 \times 10^{-3} \cot \varphi$. At this time, $\boldsymbol{Q}(\boldsymbol{x})$ is positive definite, and the control system is asymptotically stable at the origin. The load slightly changes with the input impedance of the train when the EMUs pass uphill/downhill and pass the neutral section or when there is a short external disturbance. Under the influence of SOGI-PLL, $\varphi$ slightly deviates from the original angle at this time, which may increase $R_{2}$ in the downhill range of $\cot \varphi\left[-R_{2}, 0\right]$. The neural network can be iterated to further reduce the gradient value, ensuring that it is still less than $4.712 \times 10^{-3} \cot \varphi$. Thus, (20) is established, which ensures $\boldsymbol{Q}(\boldsymbol{x})$ to be positive definite. The control system is asymptotically stable at the origin.

The convergence of the TD3 algorithm used in this paper has been thoroughly discussed and proven [27]. This paper tests the effectiveness under various conditions, under which

\begin{figure}[H]
\begin{center}
  \includegraphics[alt={},max width=\columnwidth]{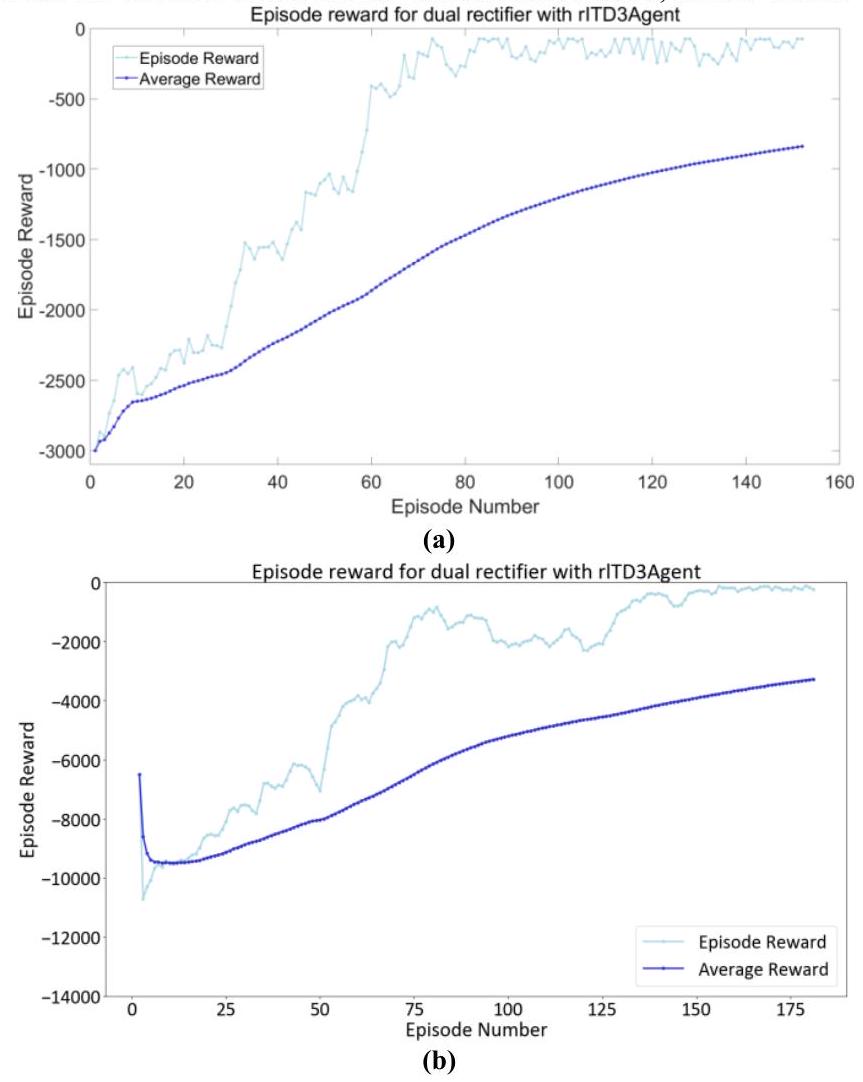}
\caption{Episode reward. (a) Stable operating condition; (b) Uphill and downhill + neutral section passing.}
\end{center}
\end{figure}

the DRL control performance is good. The corresponding DRL convergence curve is shown in Fig. 20. Therefore, the DRL approach in this paper is absolutely convergent, and there is no problem of non-convergence.

For the robustness analysis of the DRL control proposed in this paper, verification can be considered under extreme operating conditions. As can be seen from the previous discussion, the factors affecting the control effect of the\\
rectifier mainly come from the fluctuations in the front-end traction network voltage and changes in the back-end load. Therefore, extreme fluctuations in the traction network voltage and extreme changes in the load can be designed to verify robustness. According to China's Railway Technical Management Regulations, the traction network voltage fluctuation range for EMUs should be within 19 kV to 29 kV . For changes in traction load, factors such as train speed changes and uphill/downhill slopes can be considered. Currently, the maximum test speed of the CRH5 EMU is 275 $\mathrm{km} / \mathrm{h}$, and the maximum gradient is $30 \%$.

\section*{1) Extreme traction network voltage fluctuations}
As shown in Fig. 21, the traction network voltage is set to drop from the rated value of 27.5 kV to 19 kV at 3 seconds, rise from 19 kV to the rated value of 27.5 kV at 6 seconds, and rise from the rated value of 27.5 kV to 29 kV at 8 seconds. Meanwhile, the EMU speed is maintained at $200 \mathrm{~km} / \mathrm{h}$ with uphill and downhill processes.

\begin{figure}[H]
\begin{center}
  \includegraphics[alt={},max width=\columnwidth]{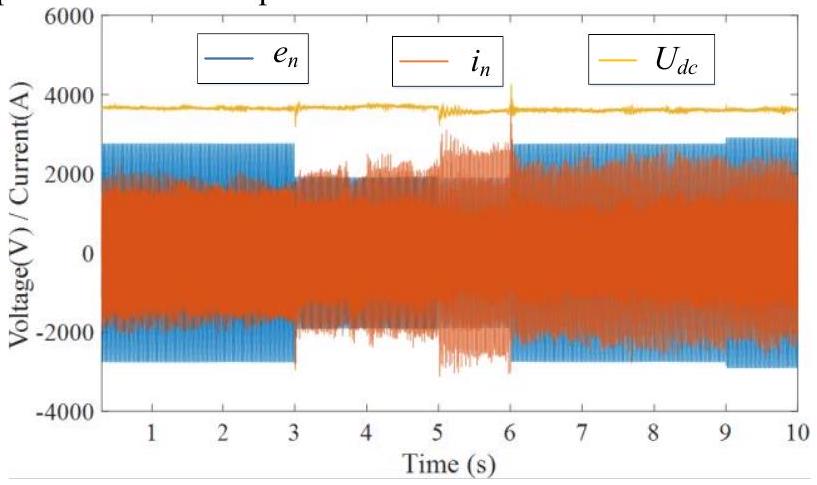}
\caption{DRL control waveforms under extreme traction network voltage fluctuations.}
\end{center}
\end{figure}

\section*{2) Extreme load changes}
As shown in Fig. 22, it is set that from 0 to 5 seconds, the EMU accelerates from rest to $275 \mathrm{~km} / \mathrm{h}$ while climbing a $30 \%$ o gradient, and from 5 to 10 seconds, it decelerates from 275 $\mathrm{km} / \mathrm{h}$ to rest while descending a $30 \%$ gradient.

\begin{figure}[H]
\begin{center}
  \includegraphics[alt={},max width=\columnwidth]{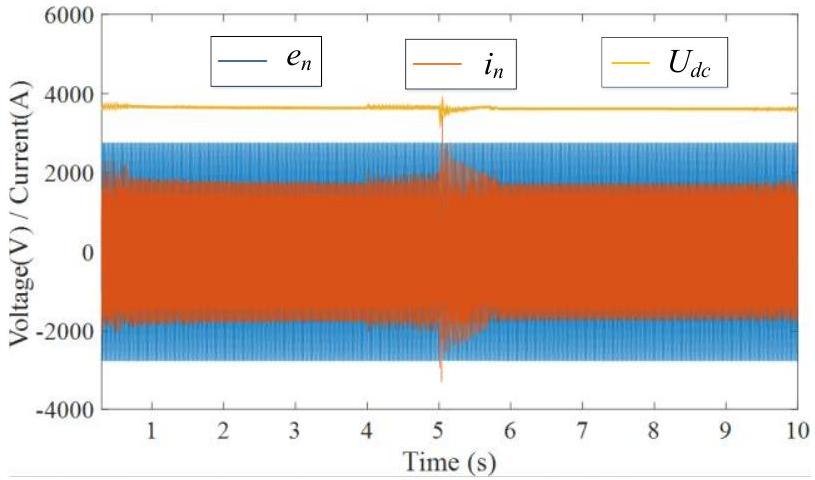}
\caption{DRL control waveforms under extreme load changes.}
\end{center}
\end{figure}

The sensitivity refers to the influence of the parameter change rate on system stability, and the corresponding values can usually be calculated by formulas. Assuming the IGBTs are ideal, in the rectifier system, and the transformers and support capacitors in actual EMUs are fixedly packaged, and their parameters generally do not change. The system parameters are set to change once every 1 second, as shown in TABLE VI. Meanwhile, to distinguish the sensitivity under\\
different parameter changes, when one parameter is varied, the other parameter remains unchanged. However, since the DRL control adopted in this paper makes it difficult to calculate numerical values, the results are reflected by simulation results, as shown in Fig. 23.

\begin{table}[H]
\begin{center}
\caption{The Change of System Parameters Per Second}
\begin{tabular}{|l|l|l|l|l|l|}
\hline
System parameters & 1s & 2s & 3s & 4s & 5s \\
\hline
Traction network voltage (kV) & 23.5 & 21.7 & 25.3 & 27.9 & 19.8 \\
\hline
Vehicle-side load (kW) & 1372 & 1620 & -158.4 & 1854 & -774 \\
\hline
System parameters & 6s & 7s & 8s & 9s & 10s \\
\hline
Traction network voltage (kV) & 22.1 & 26.4 & 24.0 & 28.6 & 20.2 \\
\hline
Vehicle-side load (kW) & 1566 & -900 & 1782 & -1152 & 2124 \\
\hline
\end{tabular}
\end{center}
\end{table}

\begin{figure}[H]
\begin{center}
  \includegraphics[alt={},max width=\columnwidth]{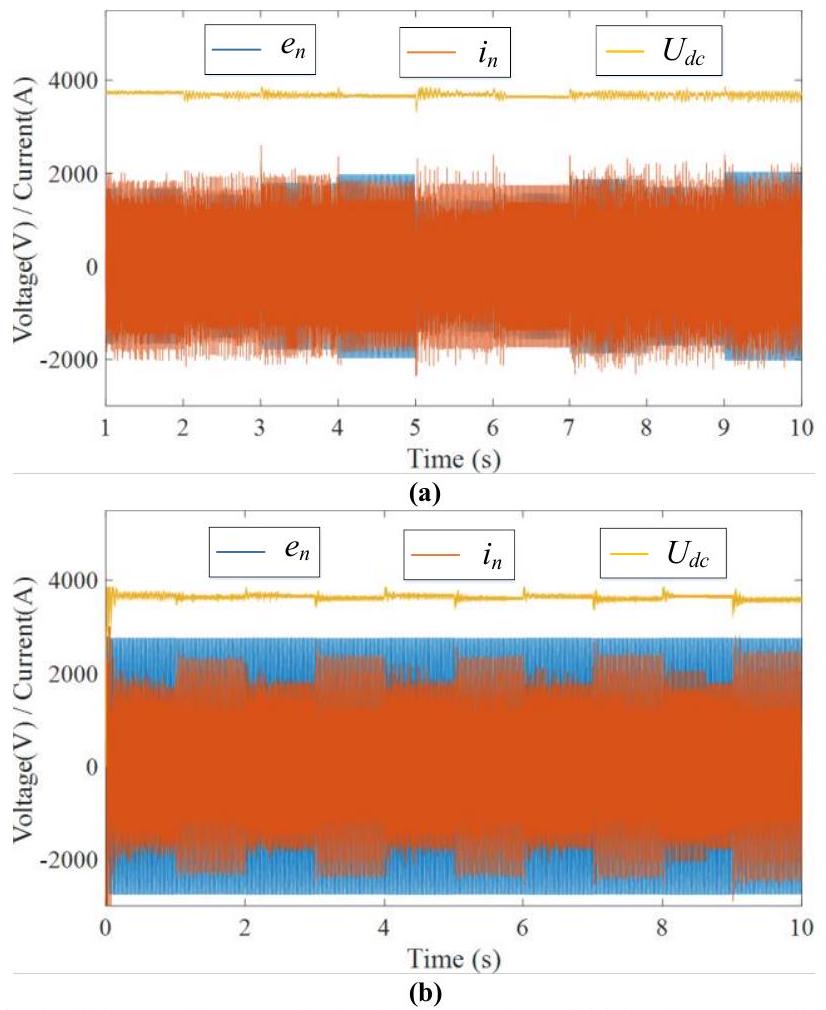}
\caption{The sensitivity analysis of the controller; (a) Traction network voltage changes (b) Vehicle-side load changes.}
\end{center}
\end{figure}

To further address the controller's generalization ability across different EMU models, which may have different electrical parameters, we conducted a model mismatch test.

In this test, the DRL agent, which was fully trained on the CRH5 parameters, was directly applied-without any retraining-to a simulation model with perturbed parameters. We perturbed the key electrical parameters, namely the leakage inductance of the onboard transformer ( $L_{n}$ ) and the DC-link capacitance ( $C_{d}$ ), by $\pm 10 \%$ to simulate the characteristics of a different EMU model.

The controller's performance under this parameter mismatch across the entire multi-working condition scenario (including\\
flat ground traction, uphill traction, and downhill regenerative braking) is shown in Fig. 24.

As shown in Fig. 24(a) and (b), the DRL agent successfully maintains the stability of the DC voltage around 3600 V . Although minor deviations in the transient response are observed, the controller demonstrates strong robustness. This test indicates that the DRL agent has not overfitted to the specific parameters of the CRH5 but has instead learned a robust control policy with good generalization capabilities.

\begin{figure}[H]
\begin{center}
  \includegraphics[alt={},max width=\columnwidth]{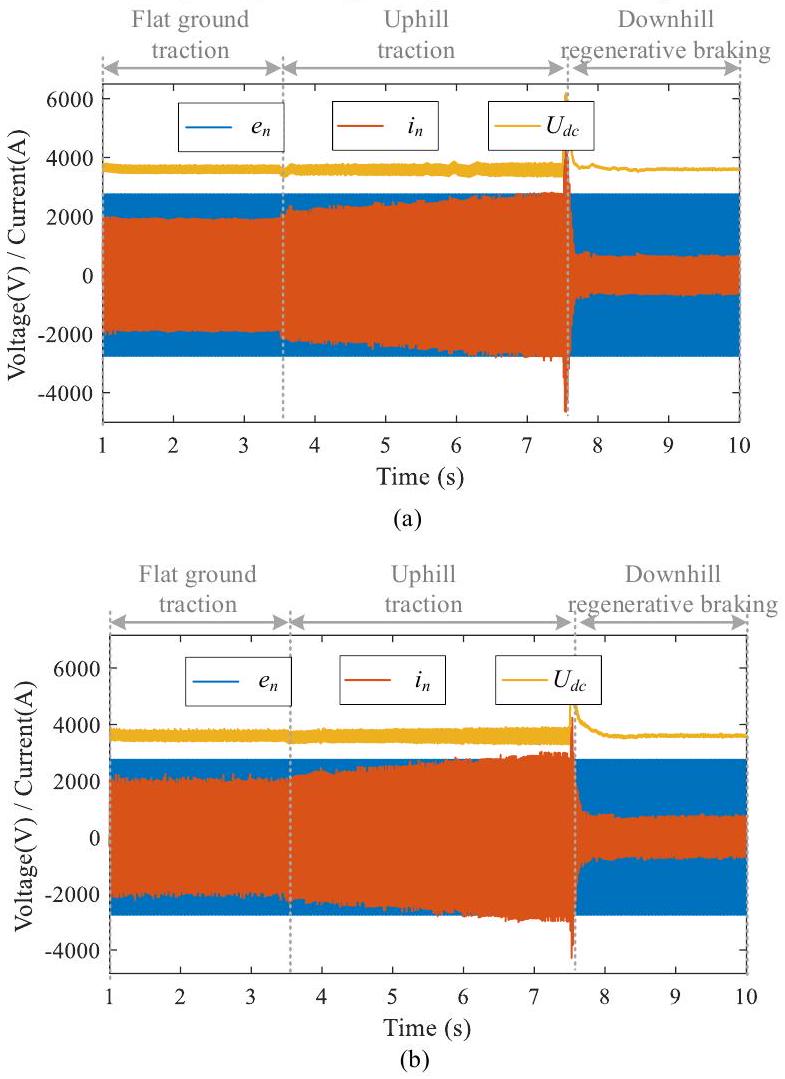}
\caption{Test of the DRL agent under model parameter mismatch. (a) Response with $+10 \%$ perturbation in $L_{n}$ and $C_{d}$. (b) Response with $-10 \%$ perturbation in $L_{n}$ and $C_{d}$.}
\end{center}
\end{figure}

According to China's Railway Technical Management Regulations and GB/T 37863.2-2021 Rail Transit-Traction Electric Drive System -Part 2: Locomotives and EMUs, the voltage limits and the switching frequency limit are as shown in TABLE VII.

\begin{table}[H]
\begin{center}
\caption{The Voltage Limits and the Switching Frequency Limit}
\begin{tabular}{|l|l|l|l|}
\hline
Indicators & Specific content & Data & Standards \\
\hline
The voltage limits & The traction network voltage fluctuation range for EMUs & $19 \mathrm{kV} \sim 29 \mathrm{kV}$ & China's Railway Technical Management Regulations \\
\hline
The switching frequency limit & The fixed switching frequency of CRH5 EMU & 250 Hz & GB/T 37863.22021 \\
\hline
\end{tabular}
\end{center}
\end{table}

We have now conducted a robustness analysis of the proposed control under extreme operating conditions, including voltage/current limits. The results show that the DRL controller exhibits good robustnes. Additionally, to\\
ensure that the DRL control model can converge in all scenarios, it is necessary to train as many situations as possible and maintain the stability of the DC voltage. This paper only tests the scenarios of EMU uphill/downhill and phase transition, and other scenarios can be further considered in the follow-up research.

\section*{IV. Semi-Physical Experiment Verification}
The computational overhead of DRL-based control refers to the computational resources and time costs consumed by the system, mainly stemming from algorithmic complexity, data interaction, and task integration, including: (1) Neural Network Training and Inference - training involves optimizing deep networks (e.g., via backpropagation), with highdimensional states or complex architectures exponentially increasing computational load and requiring massive sample storage/processing, while inference demands high-frequency real-time forward propagation to avoid control delays; (2) Environmental Interaction and Sampling - complex simulation models incur high single-interaction costs, and real-world hardware limitations reduce sampling frequency, prolonging training cycles; (3) Algorithm Optimization and Hyperparameter Tuning - hyperparameter sensitivity necessitates extensive repetitive training, further increasing computational burdens.

To further verify the control effect of the proposed method on the dual rectifiers under various working conditions and

\begin{figure}[H]
\begin{center}
  \includegraphics[alt={},max width=\columnwidth]{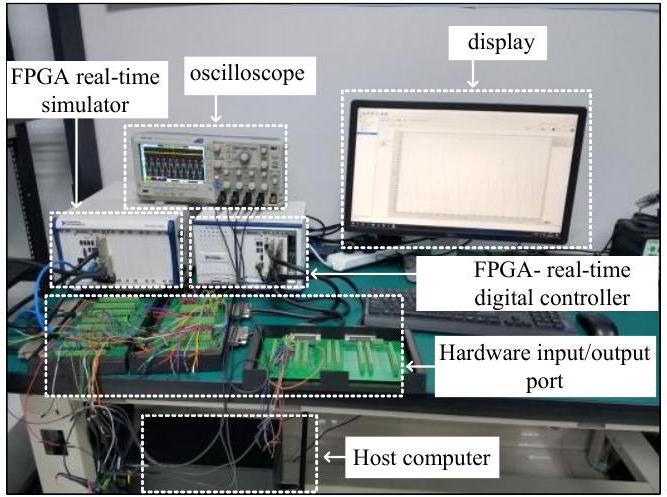}
\caption{(a)}
\end{center}
\end{figure}

\begin{figure}[H]
\begin{center}
  \includegraphics[alt={},max width=\columnwidth]{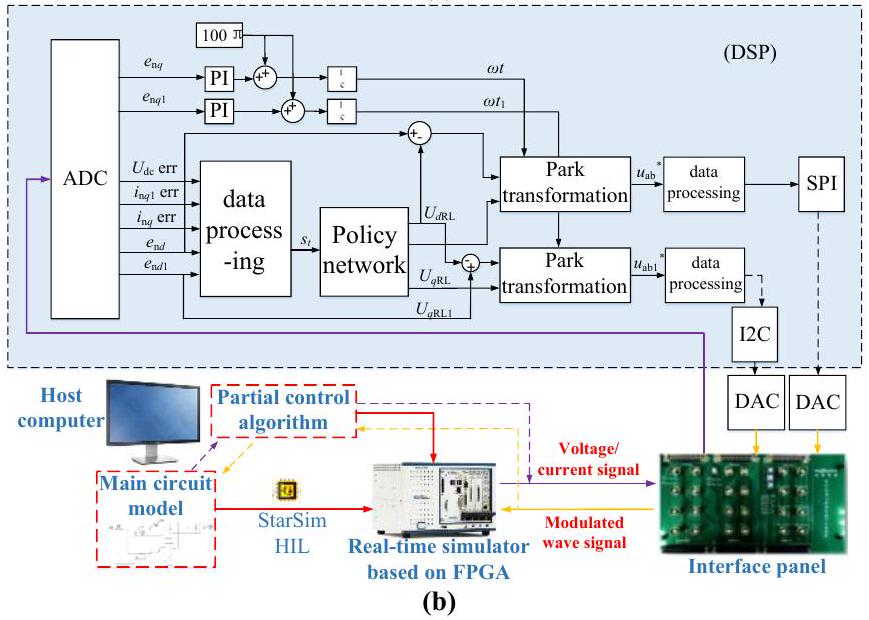}
\caption{(a) HIL hardware-in-loop simulation platform; (b) Schematic of the overall design of the experiment.}
\end{center}
\end{figure}

switching, the controller is implemented in the hardware-in-the-loop (HIL) simulation platform, as shown in Fig. 25(a). It consists of a real-time simulator (RTS) based on NI PXle-1071 FPGA, a NI PXle-1082 real-time controller (RTC) with a rapid control prototype (RCP) module, an oscilloscope, an external host, hardware input/output ports, and a power supply unit.

However, due to the software version mismatching, the experiment is special and needs to be redesigned. The overall experimental scheme using an external DSP chip is designed, as shown in Fig. 25(b).

\section*{A. Overall Design of the Experimental Scheme}
To avoid issues such as a small number of serial ports and slow data transmission rates in the serial communication scheme, this paper adopts analog-to-digital (ADC) and digital-to-analog conversion (DAC) modules with sufficient quantity and high-speed operation. Through the wiring of the interface panel, direct analog signal interaction between the DSP chip and the real-time simulator is achieved, meeting the requirements of fast control. At the same time, to avoid increasing the computational pressure on the DSP chip, the DSP chip no longer outputs $2 \times 4$ pulse signals via SPWM, but only calculates two modulation signals. Finally, two analog voltages are output to the real-time simulator through two DAC modules as shown in Fig. 26. If three control voltages are output, three DAC modules are required, which is not advisable.

\begin{figure}[H]
\begin{center}
  \includegraphics[alt={},max width=\columnwidth]{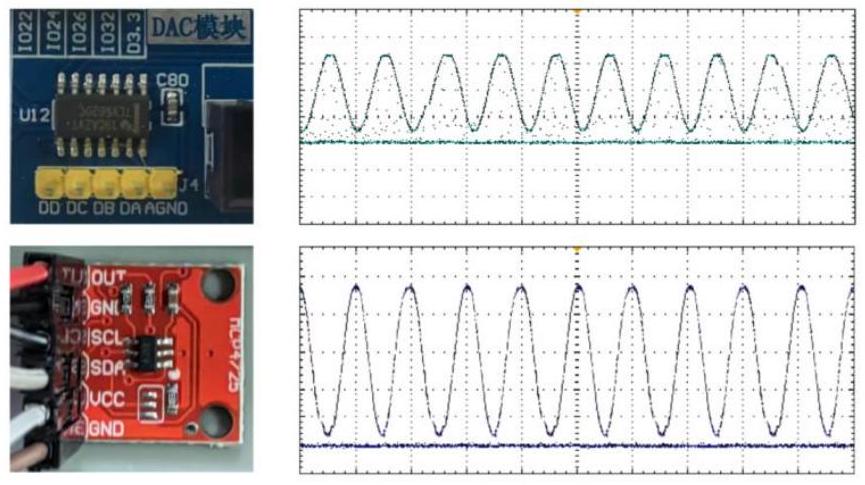}
\caption{Two DAC modules output two modulated wave.}
\end{center}
\end{figure}

Therefore, in addition to the policy network, a phase measurement and coordinate inverse transformation module should also be imported into the DSP chip, which still does not exceed the maximum computational pressure that the chip can bear. The phase measurement module uses the phase-locking method in SOGI-PLL. In addition, part of the control algorithms imported into the self-closed-loop Control Block include a voltage-current synchronization system, uphill/downhill over-voltage signal, inverter control system, four-motor current model, and modulation wave data processing module, which are mainly used for interaction with the main circuit model and external interface panel.

The scheme specifically includes: (1) ADC module design: selecting B0-B6 of channel B in the DSP, starting ADC conversion with a software-triggered source, and adopting a continuous sampling mode; (2) Importing the policy network into the DSP chip: requiring pre-installation of the deep\\
learning code generation support package and C2000 code generation support package, then using the Predict model as the carrier of the agent's policy network, generating Simulink code, and importing it into the DSP via CCS6 software; (3) (3)DAC module and hardware wiring design: TLV5620 and MCP4725 output sinusoidal modulation voltage to the interface panel, and the specific details of SCI and I2C connected to both are shown in TABLE VIII during software programming configuration, with hardware wiring as shown in Fig. 27; (4) Finally, correct data docking among the real-time simulator, DSP, and DAC is required.

\begin{table}[H]
\begin{center}
\caption{Software Programming Configuration of SPI and I2C}
\begin{tabular}{|l|l|l|l|}
\hline
Configuration & Content & I2C Configuration & Content \\
\hline
Calling Method for Slave Selection & Provided by SPI Peripheral & Slave Address Format & 7-bit \\
\hline
Clock Polarity & Triggered on Rising Edge & Repeat Mode & Enabled \\
\hline
Clock Mode & No Delay & Stop Condition & Disabled \\
\hline
Data Transmission Length & 11-bit & Data Transmission Length & 8-bit \\
\hline
\end{tabular}
\end{center}
\end{table}

\begin{figure}[H]
\begin{center}
  \includegraphics[alt={},max width=\columnwidth]{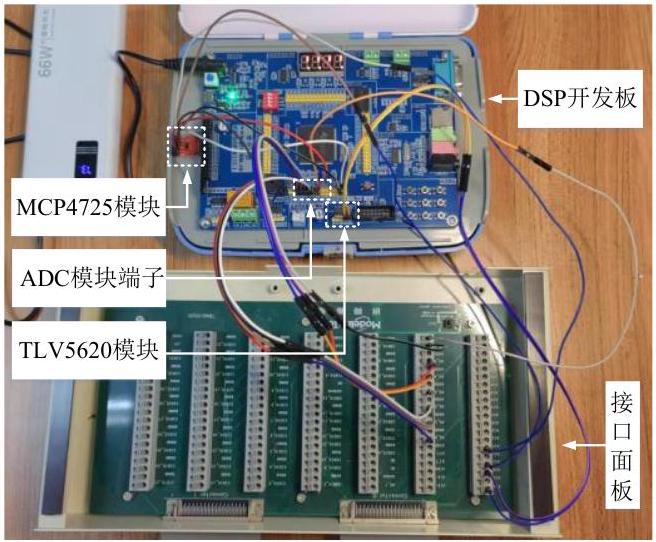}
\caption{Hardware wiring diagram.}
\end{center}
\end{figure}

In this experiment, fast control is simulated by setting different step sizes for the control system in the DSP chip and the main circuit in the simulator ( $1 \times 10^{-5} \mathrm{~s}$ and $1 \times 10^{-4} \mathrm{~s}$, respectively). For implementation on real hardware, subsequent experiments with an "online training - offline use" approach are required for verification. In practical scenarios, a DRL agent can be set up on a test train using a computer to transmit signals for controlling the rectifier. Multiple online training sessions can be conducted on a specific railway line to cover as many scenarios as possible. Finally, the trained agent can be deployed offline for practical use.

\section*{B. Experimental results of different control modes}
Fig. 28 shows the experimental waveforms of PI control and DRL-based control on the HIL simulation platform, which correspond to the situation of Fig. Fig. 10. catenary voltage $e_{\mathrm{n}}$ and catenary current fluctuate under various working conditions and switching, and the DC voltage $U_{\mathrm{dc}}$ under PI control significantly fluctuates. $U_{\mathrm{dc}}$ based on DRL control can maintain stability better. This shows that DRL-based control\\
has better control performance under various working conditions and switching.

\begin{figure}[H]
\begin{center}
  \includegraphics[alt={},max width=\columnwidth]{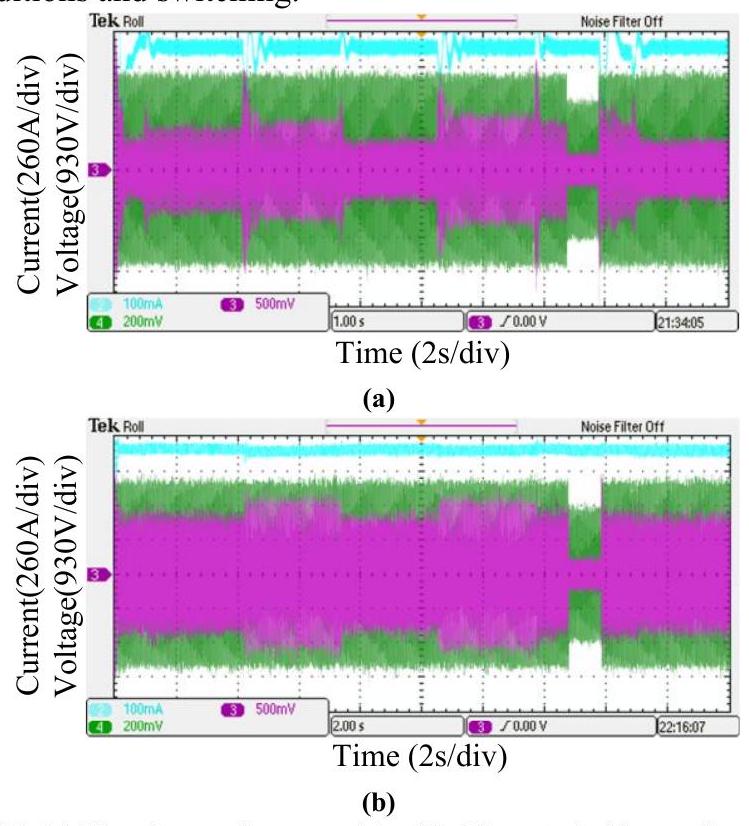}
\caption{(a) Waveforms of $\boldsymbol{u}_{\mathrm{dc}}, \boldsymbol{e}_{\mathrm{n}}$ and $\boldsymbol{i}_{\mathrm{n}}$ with PI control; (b) waveforms of $\boldsymbol{u}_{\mathrm{dc}}, \boldsymbol{e}_{\mathrm{n}}$ and $\boldsymbol{i}_{\mathrm{n}}$ with DRL control.}
\end{center}
\end{figure}

\section*{C. Experimental results of different operating processes}
In Fig. 29, the experimental waveforms of the DRL-based control applied to different operating processes of EMUs are displayed, corresponding to the two operating processes shown in Fig. 17.

\begin{figure}[H]
\begin{center}
  \includegraphics[alt={},max width=\columnwidth]{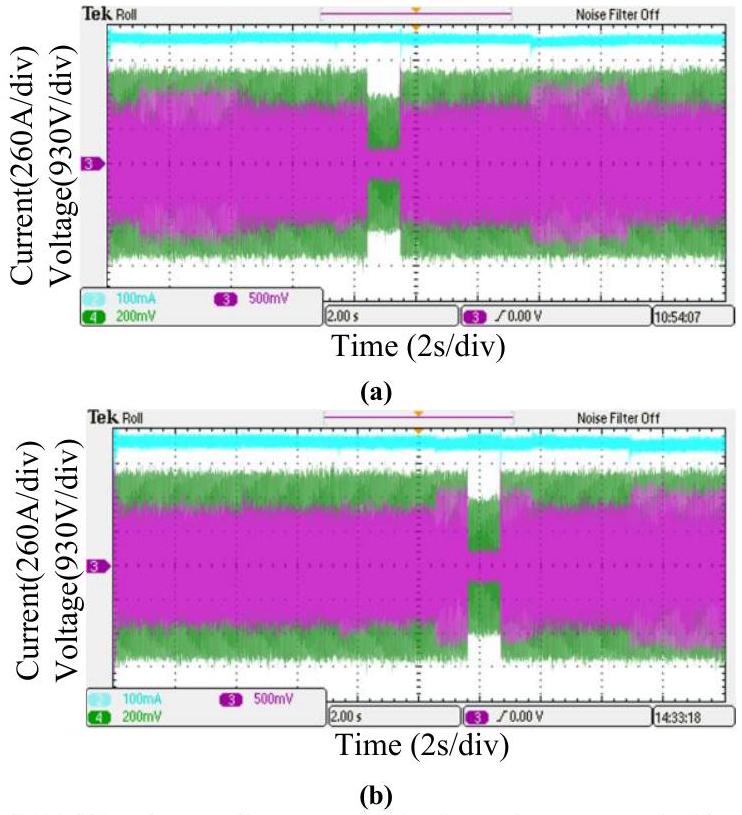}
\caption{(a) Waveforms of $\boldsymbol{u}_{\mathrm{dc}}, \boldsymbol{e}_{\mathrm{n}}$ and $\boldsymbol{i}_{\mathrm{n}}$ in Operation process 1; (b) waveforms of $\boldsymbol{u}_{\mathrm{dc}}, \boldsymbol{e}_{\mathrm{n}}$ and $\boldsymbol{i}_{\mathrm{n}}$ in Operation process 2.}
\end{center}
\end{figure}

In Fig. 29, $U_{\mathrm{dc}}$ based on DRL control can remain stable under different process conditions. It can be proved that DRLbased control has a strong generalization ability.

\section*{C. Hyperparameter tuning}
To verify the robustness of the proposed DRL control policy to the key hyperparameters, we conducted a series of sensitivity analyses.

\begin{figure}[H]
\begin{center}
  \includegraphics[alt={},max width=\columnwidth]{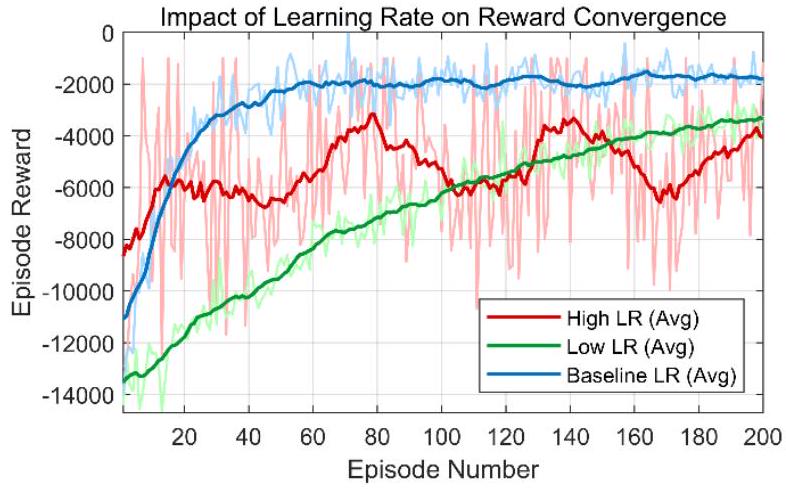}
\caption{Comparison of reward curves at different learning rates}
\end{center}
\end{figure}

(1) Learning Rate. In our experiments, the baseline learning rate was set as follows: Actor (policy network) $\operatorname{Lr}: 1 \times 10^{-3}$, Critic (value network) Lr: $1 \times 10^{-4}$. Around this baseline, we tested two additional learning rate settings: "High LR" (Actor $\mathrm{Lr}=5 \times 10^{-3}$, Critic $\mathrm{Lr}=5 \times 10^{-4}$ ) and "Low LR" (Actor $\mathrm{Lr}= 1 \times 10^{-4}$, Critic $\operatorname{Lr}=1 \times 10^{-5}$ ). All experiments used a oneepisode all-situation training method, with other hyperparameters remaining constant.

Fig. 30 illustrates the convergence curves of the "episode reward" under these three different learning rate settings. The reward curve for a high learning rate exhibits extreme instability and violent oscillations. Although the reward value rises rapidly in the early stages of training, it quickly begins to fluctuate significantly, failing to converge to a stable highreward range. This is because an excessively high learning rate leads to a massive policy update step size, ultimately causing training to diverge or converge to a very poor local optimum. The reward curve for a low learning rate is very smooth with slight variance, but the convergence speed is extremely slow. Even with the same number of training episodes, the final reward value is far lower than the baseline learning rate. An excessively low learning rate leads to slow updates to network weights, requiring the agent to spend significant time exploring for an effective policy.

\begin{figure}[H]
\begin{center}
  \includegraphics[alt={},max width=\columnwidth]{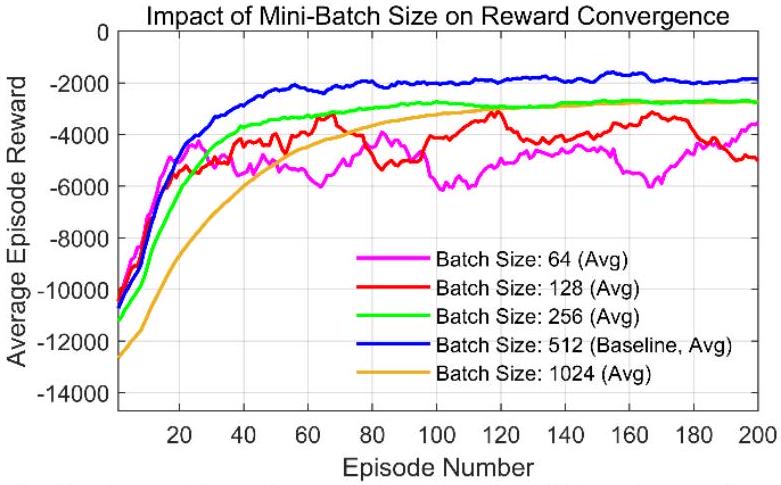}
\caption{Comparison of reward curves under different batch sizes}
\end{center}
\end{figure}

(2) Batch size: To further quantify the impact of mini-batch size on training results, we compared the final average reward after training with five different batch sizes ( $64,128,256,512$, 1024). Figure 30 shows the convergence curves of the smoothed average reward under five different batch size settings.

Fig. 31 show that the reward curves for small batches exhibit extremely high instability and violent oscillations. Although they may rise rapidly in the early stages, they quickly get trapped in poor local optima due to their extremely high variance. This is because excessively small batches lead to high-variance gradient estimates, resulting in an extremely unstable training process that fails to converge to an effective policy. In addition, the reward curves for large batches are extremely smooth, with slow convergence. This is because large batches provide very accurate (low-variance) gradient estimates, making the update process very stable. However, this stability sacrifices stochasticity. Moderate noise helps DRL explore and "escape" local optima.

\begin{figure}[H]
\begin{center}
  \includegraphics[alt={},max width=\columnwidth]{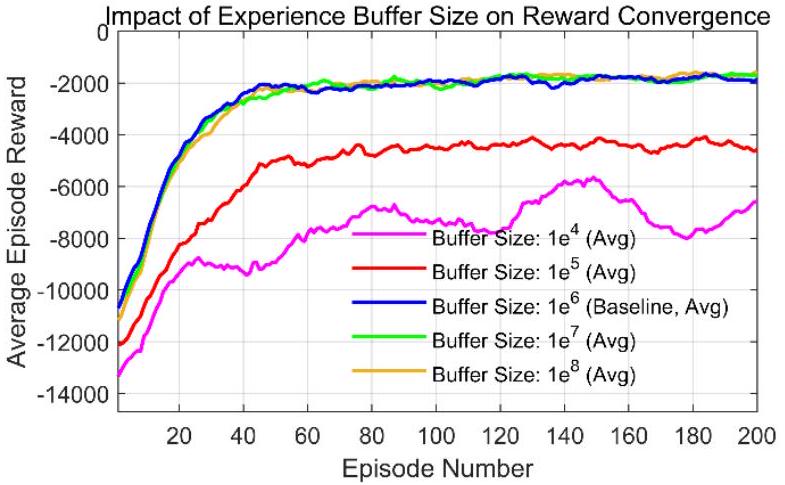}
\caption{Comparison of reward curves under different Buffer Sizes}
\end{center}
\end{figure}

(3) Buffer Size: The size of the experience replay buffer determines the agent's "memory capacity." In our experiments, the baseline buffer size was set to $1 \times 10^{6}$. We additionally tested four settings, covering a range from $10^{4}$ to $10^{8}$. Fig. 31 shows the smoothed average reward convergence curves for these five different buffer size settings.

Fig. 32 clearly shows the significant impact of buffer size on the training process. When the buffer is too small (Buffer Size $10^{4}$ or $10^{5}$ ), the reward curve converges slowly, and the final converged reward value is the worst. This is because a small buffer leads to a severe lack of diversity in experience. In our "one-episode all-situation" training, an Episode contains multiple situational transitions. A buffer that is too small will quickly fill up with new, highly relevant experiences, causing the agent to rapidly "forget" previously learned but equally essential experiences. The model exhibits optimal convergence speed and the highest final reward when the buffer size reaches $10^{6}$. When we further increase the buffer size to $10^{7}$ and $10^{8}$, the reward curve largely overlaps with that of $10^{6}$, with almost no significant performance improvement. This indicates that a size of $10^{6}$ is sufficient to\\
store diverse experiences representing the vast majority of situational transitions in our complex training environment. At this point, PER can effectively sample key transitional experiences from this, thereby learning robust and generalizable policies. Further increasing the buffer size does not bring additional performance gains but instead significantly increases memory overhead, which is not costeffective in practical engineering deployments.

\section*{V. Conclusion}
A DRL control strategy based on TD3+RS+PER is proposed, which applies AI-based data-driven control to the control of traction rectifiers. This can improve nonlinear control performance, suppress DC voltage fluctuations of high-speed trains under various operating conditions and switches, and ensure that the rectifier circuit operates close to unit power. Based on the proposed controller, simulation scenarios of EMUs running on flat ground, passing uphill/downhill sections, and passing neutral sections are built. The control performance is compared with traditional $d q$ decoupling control, and simulation experiments of the other two scenarios are conducted to verify its generalization ability. Finally, the control performance and generalization ability of the proposed controller are further verified in the HIL simulation platform. The main conclusions are listed as follows.

\section*{REFERENCES}
[1] Z. Liu, Z. Geng, S. Wu, X. Hu and Z. Zhang, "A Passivity-Based Control of Euler-Lagrange Model for Suppressing Voltage LowFrequency Oscillation in High-Speed Railway," in IEEE Transactions on Industrial Informatics, vol. 15, no. 10, pp. 5551-5560, Oct. 2019.\\[0pt]
[2] X. Meng, D. Xie, H. Lin, C. Lin, X. Ge, and Z. Liu, "DissipativityBased Multiport Stability Root-Cause Identification and Mitigation for Solid-State Transformers," IEEE Trans. Ind. Electron., early access, 2026, doi: 10.1109/TIE.2026.3658639.\\[0pt]
[3] X. Meng, Q. Zhang, Z. Liu, G. Hu, F. Liu and G. Zhang, "Multiple Vehicles and Traction Network Interaction System Stability Analysis and Oscillation Responsibility Identification," in IEEE Transactions on Power Electronics, vol. 39, no. 5, pp. 6148-6162, May 2024.\\[0pt]
[4] M. S. Sadabadi, X. Meng and Z. Liu, "Resilient and Robust Voltage Regulation in Shipboard DC Microgrids with ZIP Loads under Actuator and Parameter Uncertainties," in IEEE Transactions on Transportation Electrification, doi: 10.1109/TTE.2025.3619129.\\[0pt]
[5] Z. Liu, Q. Yan, I. A. Tasiu, Y. Zhang, K. Hu and T. Dragičević, "A Model Predictive Control Considering Parameters and System Uncertainties for Suppressing Low-Frequency Oscillations of Traction Dual Rectifiers," in IEEE Transactions on Transportation Electrification, vol. 7, no. 3, pp. 1031-1046, Sept. 2021.\\[0pt]
[6] X. Zhang, H. Bai and M. Cheng, "Improved Model Predictive Current Control With Series Structure for PMSM Drives," in IEEE Transactions on Industrial Electronics, vol. 69, no. 12, pp. 1243712446, Dec. 2022.\\[0pt]
[7] Z. Wang, Y. Liu, Z. Guan and Y. Zhang, "An Adaptive Sliding Mode Motion Control Method of Remote Operated Vehicle," in IEEE Access, vol. 9, pp. 22447-22454, 2021.\\[0pt]
[8] M. U. Hassan, M. Humayun, P. Fu, Z. Song and L. Hua, "Fuzzy Controller Using Circulating Mode for ITER Poloidal Field AC/DC Converter System," in IEEE Transactions on Plasma Science, vol. 47, no. 6, pp. 2890-2895, June 2019.\\[0pt]
[9] M. Gheisarnejad, H. Farsizadeh and M. H. Khooban, "A Novel Nonlinear Deep Reinforcement Learning Controller for DC-DC Power\\
(1) Based on the DRL control of a single rectifier, this paper can improve the DRL control of a dual rectifier. The performance indexes of DC voltage, such as overshoot, THD, stability time, and voltage fluctuation, are all improved under stable conditions, compared with PI control.\\
(2) Because of the poor control effect of the EMUs under various operating conditions, such as flat running, uphill and downhill running, and passing a neutral section, the reward shaping added in this paper can dynamically adjust the weight coefficients of each error term in the reward function according to the $U_{\mathrm{dc}}$ err. On the premise of ensuring stable DC voltage, the power factor of the unit can be guaranteed. This can not only better adapt to the jump of actions, but also suppress fluctuations of the grid voltage.\\
(3) In addition to the TD3 algorithm and Reward Shaping, Prioritized Experience Replay is added, which can effectually improve the convergence speed of the episode reward.\\
(4) After training the DRL agent in one episode that contains all cases, the generalization ability of the agent can be improved, so that it can automatically recognize various conditions and their switches. Finally, the DC voltage fluctuation can be suppressed under various working conditions. Thus, the nonlinear control can be realized to improve the robustness of the system.\\
(5) The experimental results on the HIL verify the good performance of the DRL control strategy in the dual rectifiers and its generalization ability in multi-working conditions.

Buck Converters," in IEEE Transactions on Industrial Electronics, vol. 68, no. 8, pp. 6849-6858, Aug. 2021.\\[0pt]
[10] Y. Tang et al., "Artificial Intelligence-Aided Minimum Reactive Power Control for the DAB Converter Based on Harmonic Analysis Method," in IEEE Transactions on Power Electronics, vol. 36, no. 9, pp. 97049710, Sept. 2021.\\[0pt]
[11] M. Gheisarnejad and M. H. Khooban, "An Intelligent Non-Integer PID Controller-Based Deep Reinforcement Learning: Implementation and Experimental Results," in IEEE Transactions on Industrial Electronics, vol. 68, no. 4, pp. 3609-3618, April 2021.\\[0pt]
[12] H. Wang, Z. Liu, Z. Han, Y. Wu and D. Liu, "Rapid Adaptation for Active Pantograph Control in High-Speed Railway via Deep Meta Reinforcement Learning," in IEEE Transactions on Cybernetics, vol. 54, no. 5, pp. 2811-2823, May 2024.\\[0pt]
[13] Y. Wang, S. Fang, J. Hu and D. Huang, "A Novel Active Disturbance Rejection Control of PMSM Based on Deep Reinforcement Learning for More Electric Aircraft," in IEEE Transactions on Energy Conversion, vol. 38, no. 2, pp. 1461-1470, June 2023.\\[0pt]
[14] Z. Gao, L. Ding, Q. Xiong, Z. Gong and C. Xiong, "Image Compressive Sensing Reconstruction Based on z-Score Standardized Group Sparse Representation," in IEEE Access, vol. 7, pp. 90640-90651, 2019.\\[0pt]
[15] S. Fujimoto, H.vanHoof, D. Meger, "Addressing Function Approximation Error in Actor-Critic Methods," in International Conference on Machine Learning, vol. 80, pp. 90640-90651,Jul 2018.\\[0pt]
[16] X. Meng et al., "Conversion and SISO Equivalence of Impedance Model of Single-Phase Converter in Electric Multiple Units," IEEE Trans. Transp. Electrific., vol. 9, no. 1, pp. 1363-1378, Mar. 2023.\\[0pt]
[17] Q. Zhang et al., "Modeling of Regenerative Braking Energy for Electric Multiple Units Passing Long Downhill Section," in IEEE Transactions on Transportation Electrification, vol. 8, no. 3, pp. 3742-3758, Sept. 2022.\\[0pt]
[18] Q. Xiao, P. Mattavelli, A. Khodamoradi and F. Tang, "Analysis of transforming dq impedances of different converters to a common reference frame in complex converter networks," in CES Transactions on Electrical Machines and Systems, vol. 3, no. 4, pp. 342-350, Dec. 2019.\\[0pt]
[19] J. Huang, Electric traction AC drive and control.Beijing. China Machine Press, 1998.\\[0pt]
[20] G. W. Chang, Hsin-Wei Lin and Shin-Kuan Chen, "Modeling characteristics of harmonic currents generated by high-speed railway traction drive converters," in IEEE Transactions on Power Delivery, vol. 19, no. 2, pp. 766-773, April 2004.\\[0pt]
[21] M. -T. Kuo and W. -Y. Lo, "Magnetic Components Used in the Train Pantograph to Reduce the Arcing Phenomena," in IEEE Transactions on Industry Applications, vol. 50, no. 4, pp. 2891-2899, 2014.\\[0pt]
[22] Wang H, Liu Z, Hu G, et al. Offline meta-reinforcement learning for active pantograph control in high-speed railways[J]. IEEE Transactions on Industrial Informatics, 2024, 20(8): 10669-10679.\\[0pt]
[23] OTTERLO M V,WIERING M. Reinforcement learning and markov decision processes.Berlin,Heidelberg:Springer,2012:3-42.\\[0pt]
[24] T. Okudo and S. Yamada, "Learning Potential in Subgoal-Based Reward Shaping," in IEEE Access, vol. 11, pp. 17116-17137, 2023.\\[0pt]
[25] Schaul, T., et al,"Prioritized experience replay,"in Proceedings of the 32nd International Conference on Machine Learning and Systems,2015.\\[0pt]
[26] A. Polyakov, D. Efimov, W. Perruquetti and J. -P. Richard, "Implicit Lyapunov-Krasovski Functionals for Stability Analysis and Control Design of Time-Delay Systems," in IEEE Transactions on Automatic Control, vol. 60, no. 12, pp. 3344-3349, Dec. 2015.\\[0pt]
[27] S. Fujimoto, H. Hoof, and D. Meger, "Addressing function approximation error in actor-critic methods," in International Conference on Machine Learning, PMLR, 2018, pp. 1587-1596.\\[0pt]
[28] S. Yang et al., "Learning From Human Educational Wisdom: A Student-Centered Knowledge Distillation Method," in IEEE

Transactions on Pattern Analysis and Machine Intelligence, vol. 46, no. 6, pp. 4188-4205, 2024.\\[0pt]
[29] H. Wang, Y. Song, H. Yang and Z. Liu, "Generalized Koopman Neural Operator for Data-driven Modelling of Electric Railway Pantographcatenary Systems," in IEEE Transactions on Transportation Electrification, doi: 10.1109/TTE.2025.3609347.\\[0pt]
[30] H. Yang et al., "An Electrified Railway Catenary Component Anomaly Detection Frame Based on Invariant Normal Region Prototype with Segment Anything Model," in IEEE Transactions on Transportation Electrification, doi: 10.1109/TTE.2025.3628607.\\[0pt]
[31] J. Yan, Y. Cheng, F. Zhang, et al. Research on multimodal techniques for arc detection in railway systems with limited data. Structural Health Monitoring[J], 2025, 14759217251336797.\\[0pt]
[32] Duan F, Wang H, Yang H, et al. MRFM-IFCOS: An Anchor-Free Interactive Detector Based on Multireceptive Field Mamba for Detecting Catenary Support Components[J]. IEEE Transactions on Instrumentation and Measurement, 2025, 74: 1-15.\\[0pt]
[33] X. Wang, Y. Song, H. Yang, H. Wang, B. Lu, Z. Liu, A time-frequency dual-domain deep learning approach for high-speed pantographcatenary dynamic performance prediction, Mechanical Systems and Signal Processing, 238 (2025) 113258.